\documentclass[11pt,a4paper]{article}
\pdfoutput=1
\usepackage{jheppub}
\usepackage{tikz}
\usepackage{subcaption}
\DeclareMathOperator{\Tr}{Tr}

\newcommand{\ri}{\mathrm{i}}

\newcommand{\cob}{\delta}

\newcommand{\ve}{\varepsilon}

\newcommand{\hf}{\frac{1}{2}}
\newcommand{\qu}{\frac{1}{4}}
\newcommand{\til}[1]{\widetilde{#1}}

\newcommand{\Si}{\Sigma}

\newcommand{\del}{\partial}

\newcommand{\lap}{\Delta}
\newcommand{\bra}{\langle}
\newcommand{\ket}{\rangle}

\newcommand{\h}[1]{\widehat{#1}}
\newcommand{\bt}{\beta}
\newcommand{\ga}{\gamma}

\newcommand{\al}{\alpha}

\newcommand{\Om}{\Omega}
\newcommand{\rt}[1]{\sqrt{#1}}
\newcommand{\cO}{\mathcal{O}}
\newcommand{\cZ}{\mathcal{Z}}
\newcommand{\cF}{\mathcal{F}}

\newcommand{\cD}{\mathcal{D}}
\newcommand{\cH}{\mathcal{H}}

\newcommand{\cB}{\mathcal{B}}
\newcommand{\cA}{\mathcal{A}}

\newcommand{\cM}{\mathcal{M}}
\newcommand{\cN}{\mathcal{N}}
\newcommand{\grp}[1]{\mathrm{#1}}
\newcommand{\bbZ}{{\mathbb Z}}
\newcommand{\gs}{g_{\rm s}}
\newcommand{\tilt}{\tilde{t}}
\newcommand{\tm}{{\widetilde{m}}}
\newcommand{\tZ}{\widetilde{Z}}
\newcommand{\cC}{\mathcal{C}}
\newcommand{\cT}{\mathcal{T}}
\newcommand{\vL}{L}
\newcommand{\univ}{{\rm u}}
\newcommand{\nonuniv}{{\rm nu}}
\newcommand{\MMHH}{\mathbb{HH}}
\begin{document}

\title{A proof of loop equations in 2d topological gravity}

\author[a]{Kazumi Okuyama}
\author[b]{and Kazuhiro Sakai}

\affiliation[a]{Department of Physics, Shinshu University,\\
3-1-1 Asahi, Matsumoto 390-8621, Japan}
\affiliation[b]{Institute of Physics, Meiji Gakuin University,\\
1518 Kamikurata-cho, Totsuka-ku, Yokohama 244-8539, Japan}

\emailAdd{kazumi@azusa.shinshu-u.ac.jp, kzhrsakai@gmail.com}

\abstract{
We study multi-boundary correlators in 2d Witten-Kontsevich
topological gravity. We present a proof of the loop equations
obeyed by the correlators. While the loop equations were
derived a long time ago, our proof is fully explicit
in the presence of general couplings $t_k$.
We clarify all the details, in particular the treatment of
the genus zero part of the one-boundary correlator.
The loop equations are verified by several new examples,
including the correlators of Jackiw-Teitelboim gravity
in the genus expansion and the exact correlators in the Airy case.
We also discuss the free boson/fermion
representation of the correlators
and compare it with the formulation of Marolf and Maxfield
and the string field theory of Ishibashi and Kawai.
We find similarities but also some differences.
}

\maketitle

%%%%%%%%%%%%%%%%%%%%%%%%%%%%%%%%%%%%%%%%%%%%%%%%%%%%%%%%%%%%%%%%%%%%%%%%
\section{Introduction}\label{sec:introduction}
%%%%%%%%%%%%%%%%%%%%%%%%%%%%%%%%%%%%%%%%%%%%%%%%%%%%%%%%%%%%%%%%%%%%%%%%

Two dimensional gravity coupled to a conformal matter
is a useful toy model for the study of quantum gravity.
When coupled to a particular conformal matter,
it is exactly solvable via a double-scaled
matrix model \cite{Gross:1989vs,Douglas:1989ve,Brezin:1990rb}.
For instance, 2d gravity coupled to the $(2,2k-1)$ minimal model
is described by a hermitian one-matrix model at a multi-critical point,
and it corresponds to a particular background
of Witten-Kontsevich topological gravity 
\cite{Witten:1990hr,Kontsevich:1992ti} where
the first $k$ couplings $t_n~(n\leq k)$ are turned on in a certain manner.

Recently, Saad, Shenker and Stanford showed that
Jackiw-Teitelboim (JT) gravity 
\cite{Jackiw:1984je,Teitelboim:1983ux} is also described by a certain
double-scaled matrix model \cite{Saad:2019lba}, which enables us 
to study the holographic duality
in a solvable model of 2d gravity.
It is shown in \cite{Mulase:2006baa,Dijkgraaf:2018vnm,Okuyama:2019xbv} 
that the matrix model of JT gravity is a special case of
Witten-Kontsevich topological gravity
where the infinitely many couplings $t_n$ are turned on in a specific way
(see \eqref{eq:JT-tk}).\footnote{
JT gravity can also be viewed as the $p\to\infty$ limit
of the $(2,p)$ minimal model coupled to 2d gravity \cite{Saad:2019lba,Mertens:2020hbs}.
}
This connection between JT gravity and matrix model puts the old story
of 2d gravity into the modern perspective of holography. 
In the matrix model of JT gravity \cite{Saad:2019lba}, the Hamiltonian $H$
of the boundary theory becomes the random matrix and the 
path integral of JT gravity on the asymptotically
AdS spacetime with boundary lengths
$\bt_i~(i=1,\ldots,n)$ corresponds to the ensemble average of the
partition
function $Z(\bt)=\Tr e^{-\bt H}$ over the random Hamiltonian $H$
\begin{equation}
\begin{aligned}
 \bra Z(\bt_1)\cdots Z(\bt_n)\ket.
\end{aligned} 
\label{eq:Zcorr}
\end{equation}

In the 2d gravity literature, such quantity \eqref{eq:Zcorr}
is known as the correlator of macroscopic loop operators $Z(\bt)$.
These correlators satisfy a set of equations
called the loop equation.
As we review in appendix \ref{sec:finiteN}, the loop equation of 
finite $N$ matrix model simply follows from the Schwinger-Dyson equation for the
matrix integral.
After taking the double-scaling limit of the loop equation, one can derive the 
Virasoro constraint obeyed by the partition function of the
hermitian one-matrix model \cite{Fukuma:1990jw,Dijkgraaf:1990rs}.
As shown in \cite{Itzykson:1992ya},
the same Virasoro constraint can be derived from
the Kontsevich matrix model
\cite{Kontsevich:1992ti} as well. 

The loop equation in the double-scaled matrix model is customarily
written in terms of the resolvent, which is related to the 
macroscopic loop operator $Z(\bt)$ by the Laplace transformation.
The genus expansion of the resolvent can be computed systematically
by using the topological recursion,
which basically follows from the loop equation \cite{Eynard:2007kz}.

In this paper, we will write down the loop equation
for the correlator of macroscopic loop operators $Z(\bt)$ in
Witten-Kontsevich topological gravity for the arbitrary
background $\{t_k\}$.
In particular, we will elaborate on the treatment of the genus-zero part of the one-point
function $\bra Z(\bt)\ket$, i.e.~the disk amplitude, and 
prove the loop equation in the case 
with non-zero genus-zero part $u_0$ of the specific heat $u=\gs^2\del_0^2F$ 
of topological gravity. 
We emphasize that although the loop equation of 
topological gravity was already written in the original paper \cite{Dijkgraaf:1990rs},
the details of the treatment of the disk amplitude and the $u_0\ne0$ case
for the general background have not been worked out in the literature before, 
as far as we know.
Starting from the Virasoro constraints \eqref{eq:Vircond},
we will prove the
loop equations \eqref{eq:loopfund} and \eqref{eq:loopgen} for
Witten-Kontsevich topological gravity
with arbitrary couplings $\{t_k\}$.

Our loop equation involves the effective potential $V_{\text{eff}}(\xi)$
whose explicit form was 
recently obtained for the general background with $u_0\ne0$ \cite{Okuyama:2020ncd}.
The effective potential $V_{\text{eff}}(\xi)$ is defined by the
leading term of the genus expansion of the Baker-Akhiezer function
$\psi(\xi)= e^{-\frac{1}{2\gs}V_{\text{eff}}(\xi)+\cO(\gs^0)}$
and it is related to the genus-zero part of the
eigenvalue density by
\begin{equation}
\begin{aligned}
 \rho_0(E)=\frac{1}{2\pi\gs}\text{Im}V_{\text{eff}}'(-E+\ri0).
\end{aligned} 
\end{equation}
The $u_0\ne0$ case is of particular interest since the 
edge of the spectrum of $\rho_0(E)$ is shifted from
$E=0$ to a non-zero threshold energy $E_0=-u_0$. 
Near $E=E_0$, $\rho_0(E)$ generically behaves as
\begin{equation}
\begin{aligned}
 \rho_0(E)\sim\rt{E-E_0}.
\end{aligned} 
\end{equation}
This shift of threshold energy played an important role in
the recent study of
JT gravity with conical defects \cite{Maxfield:2020ale,Witten:2020wvy}.

We revisit the loop equation in topological gravity
partly because we are motivated by
the recent discussion of the 
null state originating from the 
diffeomorphism invariance of the gravitational path integral
\cite{Marolf:2020xie}.
It has long been speculated that the Virasoro constraint of
double-scaled matrix model represents the diffeomorphism invariance of
2d gravity, and as a consequence the loop equation defines a null state.
We give a concrete expression to this argument
using the free boson/fermion representation
of the correlators which we developed in \cite{Okuyama:2020ncd}.
We find similarities and differences between our expression and that
advocated in \cite{Marolf:2020xie}.
We will give a cautionary remark
on the naive application of the sewing operation in the gravitational
path integral, which might be related to
the differences 
between the two expressions.
Our expression of loop equation has a close connection to the
closed string field theory of non-critical strings developed by 
Ishibashi and Kawai \cite{Ishibashi:1993nq,Ishibashi:1995np} 
but the details are slightly different.

This paper is organized as follows.
In section \ref{sec:multi},
we summarize the definitions and basic properties of
the multi-boundary correlators in Witten-Kontsevich topological gravity.
We also introduce a continuum analog of the Virasoro operators
and recall the explicit form of the effective potential.
In section \ref{sec:proof},
we give a precise description and a proof of
the general loop equations. We also make an interpretation of
the absence of the disk amplitude.
The loop equations are then verified
in the JT gravity and the Airy cases
as well as in the most general case.
In section \ref{sec:relation}, we first recall that the 
Virasoro constraint and the loop equation can be 
nicely expressed in the free boson/fermion 
representation of the Witten-Kontsevich $\tau$-function.
Then we comment on the difference between our result
and the approach of Marolf and Maxfield in \cite{Marolf:2020xie}.
We also comment on the similarity and difference between our result and the
closed string field theory of Ishibashi and Kawai 
\cite{Ishibashi:1993nq,Ishibashi:1995np}.
Finally we conclude in section \ref{sec:conclusion}.
Some of the details of the proof of \eqref{eq:loopfund} and \eqref{eq:loopgen}
in the main text are relegated to the appendices \ref{sec:LB}
and \ref{sec:prooftdel}.
In appendices \ref{sec:finiteN} and \ref{sec:V},
we review the loop equation at finite $N$ and the
cut-and-join representation of the Witten-Kontsevich $\tau$-function, respectively.

%%%%%%%%%%%%%%%%%%%%%%%%%%%%%%%%%%%%%%%%%%%%%%%%%%%%%%%%%%%%%%%%%%%%%%%%
\section{Multi-boundary correlators in Witten-Kontsevich gravity}
\label{sec:multi}
%%%%%%%%%%%%%%%%%%%%%%%%%%%%%%%%%%%%%%%%%%%%%%%%%%%%%%%%%%%%%%%%%%%%%%%%

%%%
\subsection{Generating function for intersection numbers}
%%%

In Witten-Kontsevich topological gravity
\cite{Witten:1990hr,Kontsevich:1992ti} (see
e.g.~\cite{Dijkgraaf:2018vnm} for a recent review) observables
are made up of the intersection numbers
\begin{align}
\langle\tau_{d_1}\cdots\tau_{d_n}\rangle_{g,n}
 =\int_{\overline{\cal M}_{g,n}}
  \psi_1^{d_1}\cdots\psi_n^{d_n},\qquad d_1,\ldots,d_n\in\bbZ_{\ge 0}.
\label{eq:intsec}
\end{align}
They are associated with
a closed Riemann surface $\Sigma$
of genus $g$ with $n$ marked points $p_1,\ldots,p_n$. We let
${\cal M}_{g,n}$ denote the moduli space of $\Sigma$ and
$\overline{\cal M}_{g,n}$ the Deligne-Mumford compactification
of ${\cal M}_{g,n}$. Here $\tau_{d_i}=\psi_i^{d_i}$ and $\psi_i$
is the first Chern class of the complex line bundle
over $\overline{\cal M}_{g,n}$ whose fiber
is the cotangent space to $p_i$. The generating function
for the intersection numbers is defined as
\begin{align}
F(\{t_k\})
 :=\sum_{g=0}^\infty \gs^{2g-2}F_g(\{t_k\}),\qquad
F_g(\{t_k\})
 :=\left\langle e^{\sum_{d=0}^\infty t_d\tau_d}\right\rangle_g.
\label{eq:genF}
\end{align}
Here $\gs$ is the genus counting parameter.
$F$ is uniquely determined 
either by the KdV equations with the string equation
\cite{Witten:1990hr,Kontsevich:1992ti}
or by the Virasoro constraints \cite{Fukuma:1990jw,Dijkgraaf:1990rs}.
In our previous papers \cite{Okuyama:2020ncd,Okuyama:2021cub}
we formulated a systematic method of computing
multi-boundary correlators 
based on the former conditions.
In this paper we instead investigate the implication of
the latter conditions for the multi-boundary correlators.

The Virasoro constraints are written as the highest weight conditions
\begin{align}
L_m e^F =0\qquad m\ge -1.
\label{eq:Vircond}
\end{align}
The Virasoro generators are given by
\begin{align}
\begin{aligned}
L_m
 &=\frac{1}{2}\sum_{k\ge 0}\frac{(2k+2m+1)!!}{(2k-1)!!}
   \tilt_k\partial_{k+m}
  +\frac{\gs^2}{4}\sum_{\substack{k,l\ge 0\\[.5ex] k+l=m-1}}
   (2k+1)!!(2l+1)!!\partial_k\partial_l
   \quad (m\ge 1),\\
L_0
 &=\frac{1}{16}+\frac{1}{2}\sum_{k\ge 0}(2k+1)\tilt_k\partial_k,\\
L_{-1}
 &=\frac{t_0^2}{4\gs^2}+\frac{1}{2}\sum_{k\ge 0}
   \tilt_{k+1}\partial_k,
\end{aligned}
\label{eq:Lm}
\end{align}
where
\begin{align}
\partial_k :=\frac{\partial}{\partial t_k},\qquad
\tilt_k := t_k-\delta_{k,1}.
\label{eq:tilt}
\end{align}
$L_m$ satisfy
\begin{align}
[L_m,L_n]=(m-n)L_{m+n}\qquad m,n\ge -1.
\label{eq:Vircomm}
\end{align}

For later convenience, let us introduce the Itzykson-Zuber variables
\cite{Itzykson:1992ya}
\begin{align}
I_n(v)\equiv I_n(v,\{t_k\}) =\sum_{m=0}^\infty t_{n+m}\frac{v^m}{m!}
\quad (n\ge 0).
\label{eq:defIn}
\end{align}
They satisfy
\begin{align}
\partial_v I_n(v) = I_{n+1}(v).
\label{eq:dIn}
\end{align}
Throughout this paper $I_n$ without specifying its argument
should always be understood as
\begin{align}
I_n=I_n(u_0)
\end{align}
with
\begin{align}
u_0&:=\partial_0^2 F_0.
\label{eq:defu0}
\end{align}
$F$ and multi-boundary correlators can be expressed
in terms of either $\{t_k\}$ or $\{I_n\}$.
The relation between $t_k$ and $I_n$
can also be expressed as
\begin{align}
t_k=\sum_{m=0}^\infty\frac{(-u_0)^m}{m!}I_{k+m}.
\label{eq:tinI}
\end{align}
The variables $I_n$ are useful because
$F_g\ (g\ge 2)$ are 
polynomials in $I_n\ (n\ge 2)$ and $(1-I_1)^{-1}$.
The (genus zero) string equation also takes the simple form
\begin{align}
I_0=u_0.
\label{eq:streq}
\end{align}
%

%%%
\subsection{Connected $n$-boundary correlator}
%%%

In this paper
we are interested in the $n$-boundary connected correlators
\begin{align}
Z_n(\beta_1,\ldots,\beta_n)
 =\langle Z(\beta_1)\cdots Z(\beta_n)\rangle_{\rm conn}.
\end{align}
They are given by the gravitational path integrals
over all possible connected Riemann surfaces with $n$ boundaries
(or more specifically,
$n$ macroscopic loops in the matrix model language)
of length $\beta_1,\ldots,\beta_n$.
They are generated from $F$ as \cite{Moore:1991ir}
\begin{align}
Z_n(\beta_1,\ldots,\beta_n)
 \simeq B(\beta_1)\cdots B(\beta_n)F,
\label{eq:ZninF}
\end{align}
where
\begin{align}
B(\beta)
=\frac{\gs}{\sqrt{2\pi}}\sum_{k=0}^\infty\beta^{k+\frac{1}{2}}\partial_k
\label{eq:Bop}
\end{align}
is the boundary creation operator.\footnote{As we saw
in \eqref{eq:intsec}, insertion of $\tau_d$
adds a marked point, i.e.~a puncture on the Riemann surface,
which corresponds to
a microscopic loop in the dual matrix model.
The macroscopic loop operator $Z(\beta)$ in the limit $\bt\to0$ 
is expanded in terms of 
the microscopic loop operator
$\tau_d$ as
\begin{equation}
\begin{aligned}
Z(\beta)\simeq\frac{\gs}{\sqrt{2\pi}}
 \sum_{d=0}^\infty\beta^{d+\frac{1}{2}}\tau_d.
\end{aligned} 
\end{equation}
The insertion of $\tau_d$ is represented by
the derivative $\partial_d$ when acting on the free energy $F$.
Therefore $B(\beta)$ of the form \eqref{eq:Bop} is viewed as the
operator that creates a macroscopic loop,
i.e.~a boundary on the Riemann surface.}
The symbol ``$\simeq$'' in \eqref{eq:ZninF} means that
the equality holds up to an additional non-universal part
\cite{Moore:1991ir}.
More specifically, we call
\begin{align}
Z_n^\univ(\beta_1,\ldots,\beta_n):=B(\beta_1)\cdots B(\beta_n)F
\end{align}
the universal parts of $Z_n$ and decompose $Z_n$ as
\begin{align}
Z_n(\beta_1,\ldots,\beta_n)
 =Z_n^\univ(\beta_1,\ldots,\beta_n)
 +Z_n^\nonuniv(\beta_1,\ldots,\beta_n).
\end{align}
The deviations $Z_n^\nonuniv$,
which we call the non-universal parts, appear
only for $n=1,2$ and are given by \cite{Okuyama:2020ncd}
\begin{align}
\begin{aligned}
Z_1^\nonuniv(\beta)
 &=\frac{1}{\gs}\sqrt{\frac{\beta}{2\pi}}
  \int_{-\infty}^0 dv\left(I_0(v)-v\right)e^{\beta v}\\
 &=\frac{1}{\sqrt{2\pi}\gs}
 \sum_{k=0}^\infty
 (-1)^k\beta^{-k-\frac{1}{2}}\tilt_k,\\[1ex]
Z_2^\nonuniv(\beta_1,\beta_2)
 &=\frac{1}{2\pi}\frac{\sqrt{\beta_1\beta_2}}{\beta_1+\beta_2}.
\end{aligned}
\label{eq:nonuniv}
\end{align}
It is easy to check that
\begin{align}
\begin{aligned}
B(\beta_2)Z_1^\nonuniv(\beta_1)&=Z_2^\nonuniv(\beta_1,\beta_2),\\
B(\beta_3)Z_2^\nonuniv(\beta_1,\beta_2)&=0.
\end{aligned}
\end{align}
Therefore, $Z_n$ for general $n$ can also be expressed as
\begin{align}
\begin{aligned}
Z_n(\beta_1,\ldots,\beta_n)
 &=B(\beta_1)\cdots B(\beta_{n-1})Z_1(\beta_n)\qquad n\ge 2,\\
Z_1(\beta)
 &=B(\beta)F+Z_1^\nonuniv(\beta).
\end{aligned}
\end{align}
%

%%%
\subsection{Full $n$-boundary correlator}
%%%

For our purposes it is convenient to
consider the full $n$-boundary correlators as well.
We let them be denoted by
\begin{align}
\cZ_n(\beta_1,\ldots,\beta_n)
 &=\langle Z(\beta_1)\cdots Z(\beta_n)\rangle.
\end{align}
They are given by the gravitational path integrals
over all possible Riemann surfaces,
including disconnected ones, with $n$ boundaries
of length $\beta_1,\ldots,\beta_n$.
They are expressed in terms of the connected correlators as
\begin{align}
\begin{aligned}
\cZ_1(\beta)&=Z_1(\beta),\\
\cZ_2(\beta_1,\beta_2)
 &=Z_2(\beta_1,\beta_2)+Z_1(\beta_1)Z_1(\beta_2),\\
\cZ_3(\beta_1,\beta_2,\beta_3)
 &=Z_3(\beta_1,\beta_2,\beta_3)
 +Z_2(\beta_1,\beta_2)Z_1(\beta_3)
 +Z_2(\beta_2,\beta_3)Z_1(\beta_1)
\\
 &\hspace{1em}
 +Z_2(\beta_3,\beta_1)Z_1(\beta_2)
 +Z_1(\beta_1)Z_1(\beta_2)Z_1(\beta_3).
\end{aligned}
\end{align}
In general, the relation between the full and the connected
correlators is expressed by means of the generating functionals
\begin{align}
\cZ[J] = e^{\cF[J]},
\end{align}
where
\begin{align}
\begin{aligned}
\cZ[J]&:=1+\sum_{n=1}^\infty\frac{1}{n!}
  \int d\beta_1\cdots d\beta_n J(\beta_1)\cdots J(\beta_n)
  \cZ_n(\beta_1,\ldots,\beta_n),\\
\cF[J]&:=\sum_{n=1}^\infty\frac{1}{n!}
  \int d\beta_1\cdots d\beta_n J(\beta_1)\cdots J(\beta_n)
  Z_n(\beta_1,\ldots,\beta_n).
\end{aligned}
\end{align}
Equivalently, in terms of $\cF[J]$
the full and the connected $n$-boundary correlators are
expressed respectively as
\begin{align}
\begin{aligned}
\cZ_n(\beta_1,\ldots,\beta_n)
&=\frac{\delta}{\delta J(\beta_1)}\cdots\frac{\delta}{\delta J(\beta_n)}
  e^\cF\bigg|_{J=0},\\
Z_n(\beta_1,\ldots,\beta_n)
&=\frac{\delta}{\delta J(\beta_1)}\cdots\frac{\delta}{\delta J(\beta_n)}
  \cF\bigg|_{J=0}.
\end{aligned}
\label{eq:ZfromgenF}
\end{align}

Let us now introduce the operator
\begin{align}
\begin{aligned}
\cB(\beta)
 &:=B(\beta)+Z_1^\nonuniv(\beta)
 =\frac{1}{\sqrt{2\pi}}
 \sum_{k=0}^\infty\left[
 \gs\beta^{k+\frac{1}{2}}\partial_k
 +\gs^{-1}(-1)^k\beta^{-k-\frac{1}{2}}\tilt_k\right],
\end{aligned}
\label{eq:hBcB}
\end{align}
where $B(\beta)$ and $Z_1^\nonuniv(\beta)$ are given
in \eqref{eq:Bop} and \eqref{eq:nonuniv} respectively.
The meaning of $\cB(\beta)$ is understood as follows.
Recall that for any operators $X,Y$ satisfying
$[X,[X,Y]]=[Y,[X,Y]]=0$
the Baker-Campbell-Hausdorff formula is written as
\begin{align}
e^Xe^Y=e^{X+Y+\frac{1}{2}[X,Y]}.
\end{align}
By setting
$X=\int d\beta J(\beta)\cB(\beta)$ and
$Y=-\int d\beta J(\beta)B(\beta)$
one obtains
\begin{align}
\begin{aligned}
e^{\int d\beta J(\beta)\cB(\beta)}
 &=e^{\int d\beta J(\beta)Z_1^{\nonuniv}(\beta)
   +\frac{1}{2}\left[\int d\beta_1 J(\beta_1) \cB(\beta_1),\,
                    -\int d\beta_2 J(\beta_2) B(\beta_2)\right]}
   e^{\int d\beta J(\beta)B(\beta)}\\
 &=e^{\int d\beta J(\beta)Z_1^{\nonuniv}(\beta)
   +\frac{1}{2}\int d\beta_1 d\beta_2 J(\beta_1)J(\beta_2)
    Z_2^\nonuniv(\beta_1,\beta_2)}
   e^{\int d\beta J(\beta)B(\beta)}.
\end{aligned}
\end{align}
From this one sees that
\begin{align}
\begin{aligned}
&e^{\int d\beta J(\beta)\cB(\beta)}e^F\\
 &=e^{\int d\beta J(\beta)Z_1^{\nonuniv}(\beta)
   +\frac{1}{2}\int d\beta_1 d\beta_2 J(\beta_1)J(\beta_2)
    Z_2^\nonuniv(\beta_1,\beta_2)}
   e^{\int d\beta J(\beta)B(\beta)}e^F\\
 &=e^{\int d\beta J(\beta)Z_1^{\nonuniv}(\beta)
   +\frac{1}{2}\int d\beta_1 d\beta_2 J(\beta_1)J(\beta_2)
    Z_2^\nonuniv(\beta_1,\beta_2)}
   e^{\sum_{n=0}^\infty\frac{1}{n!}\int d\beta_1\cdots d\beta_n
      J(\beta_1)\cdots J(\beta_n) B(\beta_1)\cdots B(\beta_n)F}\\
 &=e^Fe^{\cF[J]}.
\end{aligned}
\end{align}
Using \eqref{eq:ZfromgenF} one obtains
\begin{align}
\cZ_n(\beta_1,\ldots,\beta_n)
 =e^{-F}\cB(\beta_1)\cdots\cB(\beta_n)e^F.
\label{eq:genZn}
\end{align}
Therefore, $\cB(\beta)$ is interpreted
as the boundary creation operator
that generates the full correlators.

%%%
\subsection{Continuous Virasoro operator}
%%%

Let us introduce
\begin{align}
\vL(\beta):=\sum_{m=-1}^\infty\frac{\beta^{m+1}}{(m+1)!2^m}L_m.
\label{eq:Lofb}
\end{align}
$\vL(\beta)$ is viewed as the continuum analog of the Virasoro
generators $L_m$.
Indeed, it follows from \eqref{eq:Vircomm} that $\vL(\beta)$ satisfies
\begin{align}
[\vL(\beta_1),\vL(\beta_2)]=(\beta_1-\beta_2)\vL(\beta_1+\beta_2),
\end{align}
which is viewed as a continuum limit of \eqref{eq:Vircomm}.
Moreover,
$\vL$ and $\cB$ satisfy
\begin{align}
[\vL(\beta),\cB(\beta')]=-\beta'\cB(\beta+\beta').
\label{eq:LBcomm}
\end{align}
We give a proof of this relation in appendix \ref{sec:LB}.

%%%
\subsection{Effective potential}
%%%

The correlators $Z_n$ as well as $F$ are uniquely
characterized by the KdV equations.
The KdV equations are obtained as the compatibility conditions of
the Schr\"odinger equation
\begin{equation}
\left(\frac{\gs^2}{2}\del_0^2+u\right)\psi=\xi\psi
\end{equation}
with $u=\gs^2\partial_0^2 F$
and the KdV flow equations
(see \cite{Okuyama:2020vrh} for their explicit forms
in our convention). 
The Baker-Akhiezer function $\psi(\xi;\{t_k\})$
is a solution to these auxiliary linear differential equations.
It plays an important role in topological gravity.
In our previous papers
we systematically investigated the multi-boundary correlators $Z_n$
\cite{Okuyama:2020ncd, Okuyama:2021cub}
and the open free energy \cite{Okuyama:2020vrh}
using the fact that
they are expressed in terms of $\psi(\xi)$.
For instance, $Z_1(\beta)$ is related to $\psi(\xi)$ as
\begin{align}
\partial_0 Z_1(\beta)
 =\frac{\sqrt{2}}{\gs}
  \int_{-\infty}^\infty d\xi e^{\beta \xi}\psi(\xi)^2.
\end{align}

In terms of $\psi(\xi)$
the effective potential $V_{\rm eff}(\xi)$
is introduced as the leading order exponent in the small $\gs$ expansion
\begin{align}
\psi(\xi) = e^{-\frac{1}{2\gs}V_{\rm eff}(\xi)+{\cal O}(\gs^0)}.
\label{Veffdef}
\end{align}
The explicit form of $V_{\rm eff}(\xi)$
for general Witten-Kontsevich gravity was first obtained in
\cite{Okuyama:2020ncd}.
It is given by\footnote{
The effective potential in this paper is related to
that in \cite{Okuyama:2020ncd} by
$V_{\rm eff}^{\rm here}=\sqrt{2}V_{\rm eff}^{\rm there}$
with the identification $I_n=(-1)^n B_{n-1}\ (n\ge 2)$.
As explained in \cite{Okuyama:2020qpm},
generalization from the JT gravity case to the Witten-Kontsevich
case is straightforward.}
\begin{align}
V_{\rm eff}(\xi)
&=-2\sum_{n=1}^\infty
  \frac{I_n-\delta_{n,1}}{(2n+1)!!}2^{n+\frac{1}{2}}
  (\xi-u_0)^{n+\frac{1}{2}}.
\label{eq:Veff}
\end{align}
In this paper we will deal with its first derivative
\begin{align}
V_{\rm eff}'(\xi)
&=-2\sum_{n=1}^\infty
  \frac{I_n-\delta_{n,1}}{(2n-1)!!}2^{n-\frac{1}{2}}
  (\xi-u_0)^{n-\frac{1}{2}}.
\label{eq:dVeff}
\end{align}
In fact,
$V_{\rm eff}'$ will appear in the form of the differential operator
$V_{\rm eff}'(\partial_\beta)$, which consists of
half-integer powers of $(\partial_\beta-u_0)$.
The half-integer power of the differential operator $\partial_\beta$
is defined as (see e.g.~\cite{Dijkgraaf:1990rs})
\begin{align}
\partial_\beta^{k-\frac{1}{2}}\beta^{n-\frac{1}{2}}
&=\frac{\Gamma(n+\frac{1}{2})}{\Gamma(n-k+1)}\beta^{n-k}
\qquad (k,n\in\bbZ).
\label{eq:hfintdiff0}
\end{align}
It follows that
\begin{align}
\begin{aligned}
(\partial_\beta-u_0)^{k-\frac{1}{2}}e^{\beta u_0}\beta^{n-\frac{1}{2}}
=e^{\beta u_0}\partial_\beta^{k-\frac{1}{2}}\beta^{n-\frac{1}{2}}
=e^{\beta u_0}\frac{\Gamma(n+\frac{1}{2})}{\Gamma(n-k+1)}\beta^{n-k}.
\end{aligned}
\label{eq:hfintdiff0u0}
\end{align}
%

%%%
\subsection{One-boundary correlator and disk amplitude}
%%%

In this paper we will use the following decomposition
of the one-boundary correlator
\begin{align}
Z_1(\beta)=Z_1^{g\ge 1}(\beta)+Z_1^{g=0}(\beta).
\end{align}
The higher genus part is given by
\begin{align}
\begin{aligned}
Z_1^{g\ge 1}(\beta)
 &=B(\beta)F^{g\ge 1}\\
 &=B(\beta)\sum_{g=1}^\infty\gs^{2g-2}F_g.
\end{aligned}
\label{eq:Z1higher}
\end{align}
The genus zero part, i.e.~the disk amplitude
is given by \cite{Okuyama:2020ncd}
\begin{align}
Z_1^{g=0}(\beta)
 &=\frac{1}{\gs}\sqrt{\frac{\beta}{2\pi}}
   \int_{-\infty}^{u_0} dv\left(I_0(v)-v\right)e^{\beta v}.
\label{eq:Z1g0}
\end{align}
We further decompose it as
\begin{align}
Z_1^{g=0}(\beta)
 &=Z_1^\nonuniv(\beta)+Z_1^{\univ,g=0}(\beta),
\end{align}
where $Z_1^\nonuniv(\beta)$ is the non-universal part
given in \eqref{eq:nonuniv} and 
\begin{align}
Z_1^{\univ,g=0}(\beta)
  =\frac{1}{\gs}\sqrt{\frac{\beta}{2\pi}}
   \int_{0}^{u_0} dv\left(I_0(v)-v\right)e^{\beta v}
  =B(\beta)\frac{F_0}{\gs^2}
\label{eq:Z1univg0}
\end{align}
is the genus-zero universal part.

%%%%%%%%%%%%%%%%%%%%%%%%%%%%%%%%%%%%%%%%%%%%%%%%%%%%%%%%%%%%%%%%%%%%%%%%
\section{Proof of Loop equations}\label{sec:proof}
%%%%%%%%%%%%%%%%%%%%%%%%%%%%%%%%%%%%%%%%%%%%%%%%%%%%%%%%%%%%%%%%%%%%%%%%

%%%
\subsection{Loop equations}
%%%

The main purpose of this paper is to prove the loop equations
for general Witten-Kontsevich gravity.
The fundamental loop equation is written as
\begin{align}
\int_0^\beta ds\left[Z_2(s,\beta-s)
 +Z_1^{g\ge 1}(s)Z_1^{g\ge 1}(\beta-s)\right]
=\frac{1}{\gs}V_{\rm eff}'(\partial_\beta)
 Z_1^{g\ge 1}(\beta).
\label{eq:loopfund}
\end{align}
More generally, connected multi-boundary correlators satisfy
\begin{align}
\begin{aligned}
&\frac{1}{\gs}V'_{\rm eff}(\partial_\beta)
 \tZ_{n+1}(\beta,\beta_1,\ldots,\beta_n)\\
&=\int_0^\beta ds\left[Z_{n+2}(s,\beta-s,\beta_1,\ldots,\beta_n)
 +\sum_{I\subset S}
  \tZ_{|I|+1}(s;\beta_I)\tZ_{|S-I|+1}(\beta-s;\beta_{S-I})\right]\\
&\hspace{1em}
 +\sum_{j=1}^n\beta_jZ_n(\{\beta_i+\delta_{ij}\beta\}_{i=1}^n)\qquad
n\in\bbZ_{\ge 0}.
\end{aligned}
\label{eq:loopgen}
\end{align}
Here
\begin{align}
\begin{aligned}
\tZ_{n+1}
 &=\left\{\begin{array}{ll}
  Z_1^{g\ge 1}&\quad n=0\\[1ex]
  Z_{n+1}&\quad n\ge 1
	 \end{array}\right.,\qquad
\tZ_{|I|+1}(s;\beta_I)
 =\left\{\begin{array}{ll}
  Z_1^{g\ge 1}(s)&\quad |I|=0\\[1ex]
  Z_{|I|+1}(s,\beta_{i_1},\ldots,\beta_{i_{|I|}})&\quad |I|\ge 1
	 \end{array}\right.
\end{aligned}
\end{align}
with $I=\{i_1,i_2,\ldots,i_{|I|}\}$, $S=\{1,2,\ldots,n\}$
and the sum is taken for all possible subsets $I$ of $S$
including the empty set.
\eqref{eq:loopgen} includes \eqref{eq:loopfund} as the $n=0$ case.

As we reviewed in appendix~\ref{sec:finiteN},
the above loop equations have appearances naturally expected
from those for the finite-size matrix model.
Moreover, apart from the treatment of the genus-zero contribution,
\eqref{eq:loopfund} was derived
in \cite{Dijkgraaf:1990rs} for double-scaled matrix models.
To the best of our knowledge, however,
the precise treatment of the genus-zero contribution,
i.e.~the absence of the disk amplitude
as presented in \eqref{eq:loopfund} and \eqref{eq:loopgen}
has never been clearly stated in the literature.
Also, in \cite{Dijkgraaf:1990rs} the general form
of the loop equation was extrapolated
from that of the multi-critical models, but it is not so clear
how their derivation is generalized in the case of $u_0\ne 0$.
Having the explicit form \eqref{eq:dVeff} of $V_{\rm eff}'(\xi)$
obtained recently, we think it is meaningful to
revisit the derivation in a more specific manner.
In what follows we will present a rigorous, concrete proof of
\eqref{eq:loopfund} and \eqref{eq:loopgen}.
In section \ref{diskamp}
we will also remark on what the absence of the disk amplitude implies.

%%%
\subsection{Proof of fundamental loop equation}\label{prooffund}
%%%

Let us first prove \eqref{eq:loopfund}.
By plugging \eqref{eq:Lm} into \eqref{eq:Lofb}
the operator $\vL(\beta)$ is explicitly written as
\begin{align}
\begin{aligned}
\vL(\beta)
 &=\frac{\gs^2}{2}\sum_{k,l\ge 0}
   \frac{(2k+1)!!(2l+1)!!\beta^{k+l+2}}{(k+l+2)!2^{k+l+2}}
   \partial_k\partial_l\\
&\hspace{1em}
  +\sum_{\substack{m\ge -1,\ k\ge 0\\[.5ex] (m,k)\ne (-1,0)}}
   \frac{(2k+2m+1)!!\beta^{m+1}}{(2k-1)!!(m+1)!2^{m+1}}
   \tilt_k\partial_{k+m}
  +\frac{\beta}{16}+\frac{t_0^2}{2\gs^2}\\
 &=:\vL(\beta)\Big|_{\partial\partial}
  +\vL(\beta)\Big|_{\tilt\partial}
  +\frac{\beta}{16}+\frac{t_0^2}{2\gs^2}.
\end{aligned}
\label{eq:cL1}
\end{align}
Let us rewrite the above expression in terms of
the boundary creation operator $B(\beta)$ given in \eqref{eq:Bop}
instead of the derivative $\partial_k$.
The first term is immediately rewritten as
\begin{align}
\begin{aligned}
\vL(\beta)\Big|_{\partial\partial}
&=\frac{\gs^2}{2\pi}\sum_{k,l\ge 0}
  \frac{\Gamma(k+\tfrac{3}{2})\Gamma(l+\tfrac{3}{2})}{\Gamma(k+l+3)}
  \beta^{k+l+2}
  \partial_k\partial_l\\
&=\int_0^\beta ds B(s)B(\beta-s).
\end{aligned}
\label{eq:Vircond0_1}
\end{align}
As we prove in appendix \ref{sec:prooftdel},
the second term can be expressed as
\begin{align}
\vL(\beta)\Big|_{\tilt\partial}
 =-2\int_0^\beta ds Z_1^{\univ,g=0}(\beta-s)B(s)
  -\frac{1}{\gs}V'_{\rm eff}(\partial_\beta)B(\beta),
\label{eq:cLtdel}
\end{align}
where $Z_1^{\univ,g=0}(\beta)$ and $V'_{\rm eff}(\xi)$
are given in \eqref{eq:Z1univg0} and \eqref{eq:dVeff} respectively.
Thus \eqref{eq:cL1} is rewritten as
\begin{align}
\vL(\beta)
 &=\int_0^\beta ds B(s)B(\beta-s)
  -2\int_0^\beta ds Z_1^{\univ,g=0}(\beta-s)B(s)
  -\frac{1}{\gs}V'_{\rm eff}(\partial_\beta)B(\beta)
  +\frac{\beta}{16}+\frac{t_0^2}{2\gs^2}.
\label{eq:cL2}
\end{align}

The Virasoro constraints \eqref{eq:Vircond} imply that
\begin{align}
\vL(\beta)e^F=0.
\label{eq:cLon1}
\end{align}
By using \eqref{eq:cL2} this equation is written as
\begin{align}
\begin{aligned}
0=e^{-F}\vL(\beta)e^F
&=\int_0^\beta ds [B(s)B(\beta-s)F+(B(s)F)(B(\beta-s)F)]\\
&\hspace{1em}
  -2\int_0^\beta ds Z_1^{\univ,g=0}(\beta-s)B(s)F
  -\frac{1}{\gs}V'_{\rm eff}(\partial_\beta)B(\beta)F
  +\frac{\beta}{16}+\frac{t_0^2}{2\gs^2}\\
&=\int_0^\beta ds \left[Z_2^\univ(s,\beta-s)
    +Z_1^\univ(s)Z_1^\univ(\beta-s)\right]\\
&\hspace{1em}
  -2\int_0^\beta ds Z_1^{\univ,g=0}(\beta-s)Z_1^\univ(s)
  -\frac{1}{\gs}V'_{\rm eff}(\partial_\beta)Z_1^\univ(\beta)
  +\frac{\beta}{16}+\frac{t_0^2}{2\gs^2}\\
&=\int_0^\beta ds \left[Z_2(s,\beta-s)
   +Z_1^{g\ge 1}(s)Z_1^{g\ge 1}(\beta-s)\right]
  -\frac{1}{\gs}V'_{\rm eff}(\partial_\beta)Z_1^{g\ge 1}(\beta)\\
&\hspace{1em}
  -\int_0^\beta ds Z_1^{\univ,g=0}(\beta-s)Z_1^{\univ,g=0}(s)
  -\frac{1}{\gs}V'_{\rm eff}(\partial_\beta)Z_1^{\univ,g=0}(\beta)
  +\frac{t_0^2}{2\gs^2}.
\end{aligned}
\label{eq:looppre}
\end{align}
In the last step we have used
\begin{align}
Z_2=Z_2^\univ+Z_2^\nonuniv,\qquad
Z_1^\univ=Z_1^{g\ge 1}+Z_1^{\univ,g=0}
\label{eq:Z2Z1decomp}
\end{align}
and
\begin{align}
\int_0^\beta ds Z_2^\nonuniv(s,\beta-s)
 =\int_0^\beta ds\frac{\sqrt{s(\beta-s)}}{2\pi\beta}
 =\frac{\beta}{16}.
\end{align}
Expanding the last expression of \eqref{eq:looppre} in $\gs$, 
one finds that the first two terms give contributions
that are non-negative powers of $\gs$
while the last three terms are of the order of $\gs^{-2}$.
They both vanish independently
in order for \eqref{eq:looppre} to hold.
Thus we have proved \eqref{eq:loopfund}.

%%%
\subsection{Proof of general loop equation}\label{proofgen}
%%%

Let us next prove \eqref{eq:loopgen}.
For any function $f(\{t_k\})$ it follows from \eqref{eq:cL2} that
\begin{align}
\begin{aligned}
e^{-F}\vL(\beta)e^F f
 &=f e^{-F}\vL(\beta)e^F\\
 &\hspace{1em}
 +\int_0^\beta ds
  \left[B(s)B(\beta-s)+2(B(\beta-s)F)B(s)
  -2 Z_1^{\univ,g=0}(\beta-s)B(s)\right]f\\
 &\hspace{1em}
  -\frac{1}{\gs}V'_{\rm eff}(\partial_\beta)B(\beta)f\\
 &=\left[
  \int_0^\beta ds B(s)B(\beta-s)
 +2\int_0^\beta ds
    Z_1^{g\ge 1}(\beta-s)B(s)
  -\frac{1}{\gs}V'_{\rm eff}(\partial_\beta)B(\beta)\right]f.
\end{aligned}
\label{eq:ecL}
\end{align}
In the last step we have used \eqref{eq:cLon1} and
$BF=Z_1^\univ=Z_1^{g\ge 1}+Z_1^{\univ,g=0}$.

On the other hand, starting from \eqref{eq:cLon1}
and using \eqref{eq:genZn} and \eqref{eq:LBcomm} we obtain
\begin{align}
\begin{aligned}
0&=e^{-F}\cB(\beta_1)\cdots\cB(\beta_n) L(\beta) e^F\\
 &=e^{-F} L(\beta)\cB(\beta_1)\cdots\cB(\beta_n) e^F
  -e^{-F}[L(\beta),\cB(\beta_1)\cdots\cB(\beta_n)] e^F\\
 &=e^{-F} L(\beta)e^F\cZ_n(\beta_1,\ldots,\beta_n)
  +e^{-F}\sum_{k=1}^n\beta_k
   \cB(\beta_1)\cdots\cB(\beta_k+\beta)\cdots\cB(\beta_n) e^F\\
 &=\left[e^{-F} L(\beta)e^F+\cT(\beta)\right]
   \cZ_n(\beta_1,\ldots,\beta_n).
\end{aligned}
\label{eq:loopgenfull}
\end{align}
Here we have introduced a formal shift operator $\cT(\beta)$
which acts on any function $f$ of $n$-variables $\beta_1,\ldots,\beta_n$
as
\begin{align}
\cT(\beta) f(\beta_1,\ldots,\beta_n)
  :=\sum_{k=1}^n\beta_k f(\beta_1,\ldots,\beta_k+\beta,\ldots,\beta_n).
\end{align}

By using \eqref{eq:ecL} and \eqref{eq:ZfromgenF},
the equation \eqref{eq:loopgenfull} is expressed as
\begin{align}
\begin{aligned}
0&=\left[\int_0^\beta ds B(s)B(\beta-s)\right.\\
 &\hspace{2em}\left.
   {}+2\int_0^\beta ds
    Z_1^{g\ge 1}(\beta-s)B(s)
  -\frac{1}{\gs}V'_{\rm eff}(\partial_\beta)B(\beta)
   +\cT(\beta)\right]
  \frac{\delta}{\delta J(\beta_1)}\cdots\frac{\delta}{\delta J(\beta_n)}
  e^\cF\bigg|_{J=0}\\
 &=
 \frac{\delta}{\delta J(\beta_1)}\cdots\frac{\delta}{\delta J(\beta_n)}
 \cA e^\cF\bigg|_{J=0}
\end{aligned}
\label{eq:loopgenfull2}
\end{align}
with 
\begin{align}
\begin{aligned}
\cA
 &=\int_0^\beta ds B(s)B(\beta-s)\cF
  +\int_0^\beta ds (B(s)\cF)(B(\beta-s)\cF)
 \\
 &\hspace{1em}
   {}+2\int_0^\beta ds
    Z_1^{g\ge 1}(\beta-s)B(s)\cF
  -\frac{1}{\gs}V'_{\rm eff}(\partial_\beta)B(\beta)\cF
   +\cT(\beta)\cF.
\end{aligned}
\end{align}
Note that $B$ and $\cT$ in $\cA$
do not act on $e^\cF$
in the last line of \eqref{eq:loopgenfull2}.
We can rewrite \eqref{eq:loopgenfull2} as
\begin{align}
\begin{aligned}
0&=\left[
 \frac{\delta}{\delta J(\beta_1)}\cdots\frac{\delta}{\delta J(\beta_n)}
 \cA\right] e^\cF\bigg|_{J=0}\\
&\hspace{1em}
 +\sum_{I\subsetneq S}\prod_{i_k\in I}
 \left[
 \frac{\delta}{\delta J(\beta_{i_1})}\cdots
 \frac{\delta}{\delta J(\beta_{i_{|I|}})}
 \cA\right]
 \prod_{j_k\in S-I}
 \left[
 \frac{\delta}{\delta J(\beta_{j_1})}\cdots
 \frac{\delta}{\delta J(\beta_{j_{|S-I|}})}
 e^\cF\right]\Bigg|_{J=0},
\end{aligned}
\end{align}
where $I=\{i_1,i_2,\ldots,i_{|I|}\}$, $S=\{1,2,\ldots,n\}$
and the sum is taken for all possible proper subsets $I$ of $S$
including the empty set.
Based on this expression one can show by induction
(with respect to $n$) that
\begin{align}
\frac{\delta}{\delta J(\beta_1)}\cdots\frac{\delta}{\delta J(\beta_n)}
 \cA\bigg|_{J=0}=0.
\end{align}
This gives \eqref{eq:loopgen}.

%%%
\subsection{Remark on disk amplitude contribution}\label{diskamp}
%%%

For a better understanding of the structure of the loop equation
let us elaborate on the absence of the disk amplitude
in \eqref{eq:loopgen}.

To do this, let us first rewrite the genus-zero part
\eqref{eq:Z1g0} as follows.
Consider the Taylor series expansion of $\left(I_0(v)-v\right)$
about the point $v=u_0$.
By using the property \eqref{eq:dIn}
and the string equation \eqref{eq:streq}
it is expressed as
\begin{align}
I_0(v)-v=\sum_{n=1}^\infty (I_n(u_0)-\delta_{n,1})\frac{(v-u_0)^n}{n!}.
\label{eq:I0exp}
\end{align}
Plugging this into \eqref{eq:Z1g0} and evaluating the integral
one obtains
\begin{align}
Z_1^{g=0}(\beta)
 =\frac{e^{\beta u_0}}{\sqrt{2\pi}\gs}
  \sum_{n=1}^\infty (-1)^n\beta^{-n-\frac{1}{2}}(I_n-\delta_{n,1}).
\label{eq:Z1g0ter}
\end{align}
From this expression one observes that
the disk amplitude contains negative powers of $\beta$.
This is in contrast to the higher genus amplitudes,
which contain only nonnegative powers of $\beta$
\cite{Okuyama:2019xbv, Okuyama:2020ncd}
(see \eqref{eq:WK-12} in the next subsection).

Comparing the expression \eqref{eq:Z1g0ter} with \eqref{eq:dVeff}
one can regard $V_{\rm eff}'(\xi)$ as 
a formal Laplace transform of the disk amplitude
\begin{align}
-\frac{1}{2\gs} V_{\rm eff}'(\xi)
\sim
\int_0^\infty d\beta Z_1^{g=0}(\beta) e^{-\beta\xi}.
\label{eq:VeffZ1rel}
\end{align}
Of course,
this should not be viewed as a mathematically rigorous relation,
because the Laplace transform converges
only for $e^{\beta u_0}\beta^{-n-\frac{1}{2}}$
with $n<\frac{1}{2}$,
but actually $n$ is summed over positive integers.
Nevertheless, at the price of mathematical rigor
this formal relation provides us with an intuitive understanding of
the structure of the loop equations, as we see below.

Let us now come back to the loop equation \eqref{eq:loopgen}.
If we naively place $Z_1(\beta)$ instead of $Z_1^{g\ge 1}(\beta)$,
the r.h.s.~of \eqref{eq:loopgen} gets an extra contribution
\begin{align}
2\int_0^\beta ds Z_1^{g=0}(s)\tZ_{n+1}(\beta-s,\{\beta_k\}_{k=1}^n).
\label{eq:loopadd}
\end{align}
This gives rise to divergence due to the negative powers of $\beta$
as seen in \eqref{eq:Z1g0ter} and thus should not be included
in the loop equation.
However, let us 
be tolerant for a while and
attempt to evaluate it using \eqref{eq:VeffZ1rel}.
Let $\tZ_{n+1}^*$ denote the Laplace transform of $\tZ_{n+1}$:
\begin{align}
\tZ_{n+1}^*(\xi)
 &=\int_0^\infty d\beta e^{-\beta\xi}\tZ_{n+1}(\beta),\qquad
\tZ_{n+1}(\beta)
 =\frac{1}{2\pi\ri}\int_{\cC}
  d\xi e^{\beta\xi} \tZ_{n+1}^*(\xi).
\label{eq:LapZn1}
\end{align}
Here we have introduced the abbreviated notation
$\tZ_{n+1}^*(\xi)\equiv\tZ_{n+1}^*(\xi;\{\beta_k\}_{k=1}^n)$, 
$\tZ_{n+1}(\beta)\equiv\tZ_{n+1}\left(\beta,\{\beta_k\}_{k=1}^n\right)$
and the contour $\cC$ is chosen accordingly
so that the inverse Laplace transform makes sense.
Recall that the Laplace transform maps a convolution product
to an ordinary product.
Rewriting the convolution \eqref{eq:loopadd}
using \eqref{eq:VeffZ1rel} and \eqref{eq:LapZn1}
we see that
\begin{align}
\begin{aligned}
2\int_0^\beta ds Z_1^{g=0}(s)\tZ_{n+1}(\beta-s)
&\sim 2\frac{1}{2\pi\ri}\int_{\cC}
 d\xi e^{\beta\xi}
 \left(-\frac{1}{2\gs} V_{\rm eff}'(\xi)\right)
 \tZ_{n+1}^*(\xi)\\
&=-\frac{1}{\gs} V_{\rm eff}'(\partial_\beta)
 \frac{1}{2\pi\ri}\int_{\cC}
 d\xi e^{\beta\xi}
 \tZ_{n+1}^*(\xi)\\
&=-\frac{1}{\gs} V_{\rm eff}'(\partial_\beta)
 \tZ_{n+1}(\beta).
\end{aligned}
\end{align}
Being transposed to the other side of the loop equation,
this becomes precisely
what we have already had on the l.h.s.~of \eqref{eq:loopgen}!
Therefore, we can think that
the disk amplitude contribution \eqref{eq:loopadd}
is not removed by hand from the loop equation,
but rather it turns into the term involving
$V_{\rm eff}'(\partial_\beta)$
as a mathematically well-defined contribution.

%%%
\subsection{Examples: JT gravity and Airy case}\label{sec:example}
%%%

One can check the loop equation \eqref{eq:loopgen}
order by order in the genus expansion,
using the formalism of the genus expansion of
the multi-boundary correlators developed in our previous paper \cite{Okuyama:2020ncd}.
Although  
we have applied this formalism to the JT gravity case in \cite{Okuyama:2020ncd},
our formalism can be trivially generalized to 
2d topological gravity with arbitrary
background couplings $\{t_k\}$,
as explained in \cite{Okuyama:2020qpm}.
Indeed, we 
have checked that the loop equation \eqref{eq:loopgen} is satisfied 
up to the first few orders in
the genus expansion for the general background
$\{t_k\}$.
For instance, the one- and two-boundary correlators
in the genus expansion are obtained as
\cite{Okuyama:2019xbv, Okuyama:2020ncd}
\begin{align}
\begin{aligned}
Z_1^{g\ge 1}(\beta)
&=\frac{e^{\beta u_0}}{\sqrt{2\pi\beta^3}}\Biggl[
\left(\frac{I_2\beta^2}{24t^2}+\frac{\beta^3}{24t}\right)\gs\\
&\hspace{5em}
+\biggl[
\left(\frac{I_5}{1152t^4}
 +\frac{29I_3^2}{5760t^5}
 +\frac{11I_2I_4}{1440t^5}
 +\frac{5I_2^2I_3}{144t^6}
 +\frac{7I_2^4}{288t^7}\right)\beta^2\\
&\hspace{7em}
+\left(\frac{7I_2^3}{288t^6}
 +\frac{29I_2I_3}{1440t^5}
 +\frac{I_4}{384t^4}\right)\beta^3
+\left(\frac{7I_2^2}{480t^5}
 +\frac{29I_3}{5760t^4}\right)\beta^4\\
&\hspace{7em}
+\frac{29I_2\beta^5}{5760t^4}
+\frac{\beta^6}{1152t^3}
\biggr]\gs^3+{\cal O}(\gs^5)\Biggr],\\
Z_2(\beta_1,\beta_2)
&=\frac{\sqrt{\beta_1\beta_2}}{2\pi}e^{(\beta_1+\beta_2)u_0}
 \Biggl[\frac{1}{\beta_1+\beta_2}\\
&\hspace{5em}
 +\left(
  \frac{I_3}{24t^3}
 +\frac{I_2^2}{12t^4}
 +\frac{I_2(\beta_1+\beta_2)}{12t^3}
 +\frac{\beta_1^2+\beta_1\beta_2+\beta_2^2}{24t^2}
 \right)\gs^2
 +{\cal O}(\gs^4)
\Biggr],
\end{aligned}
\label{eq:WK-12}
\end{align}
where
\begin{align}
t=1-I_1.
\end{align}
Substituting these into \eqref{eq:loopfund}
and using the formula \eqref{eq:hfintdiff0u0}
one can check that the both sides of the equation give
\begin{align}
\frac{\beta}{16}
+\left[
 \left(\frac{I_3}{384t^3}
  +\frac{49I_2^2}{9216t^4}\right)\beta^2
 +\frac{49I_2}{9216t^3}\beta^3
 +\frac{35}{16384t^2}\beta^4
 \right]\gs^2+{\cal O}(\gs^4).
\end{align}
This confirms \eqref{eq:loopfund}, which is \eqref{eq:loopgen} for $n=0$,
up to this order.

JT gravity is a special case of topological gravity with
infinitely many couplings turned on in a specific way \cite{Mulase:2006baa,Dijkgraaf:2018vnm,Okuyama:2019xbv}
\begin{equation}
\begin{aligned}
 t_0=t_1=0,\qquad
t_k=\frac{(-1)^k}{(k-1)!}\quad(k\geq2).
\end{aligned} 
\label{eq:JT-tk}
\end{equation}
In this case
\begin{align}
u_0=I_1=0,\qquad I_k=\frac{(-1)^k}{(k-1)!}\quad(k\geq2)
\label{eq:u0I_JT}
\end{align}
and $V_{\text{eff}}'(\xi)$
in \eqref{eq:dVeff} is given by
\begin{equation}
\begin{aligned}
 V_{\text{eff}}'(\xi)=\rt{2}\sin(2\rt{\xi}).
\end{aligned} 
\label{eq:VeffJT}
\end{equation}
In JT gravity the one- and two-boundary correlators at the first few orders
in the genus expansion are given by
\cite{Saad:2019lba,Okuyama:2019xbv,Okuyama:2020ncd}\footnote{In 
\eqref{eq:JT-12} we have set the asymptotic value $\ga$
of the dilaton
as $\ga=\frac{1}{2\pi^2}$.
$S_0$ in \cite{Saad:2019lba} is related to $\gs$ by
$\gs=(2\pi^2)^{\frac{3}{2}}e^{-S_0}$ \cite{Okuyama:2019xbv,Okuyama:2020ncd}.}
\begin{equation}
\begin{aligned}
 Z_1^{g\ge 0}(\bt)&=\frac{1}{\rt{2\pi\bt^3}}\Biggl[
\frac{\bt^2+\bt^3}{24}\gs\\
&\hskip20mm+\left(\frac{29\bt^2}{3072}+\frac{169\bt^3}{11520}
+\frac{139\bt^4}{11520}+\frac{29\bt^5}{5760}+\frac{\bt^6}{1152}
\right)\gs^3+\cO(\gs^5)\Biggr],\\
Z_2(\bt_1,\bt_2)&=\frac{\rt{\bt_1\bt_2}}{2\pi}
\left[\frac{1}{\bt_1+\bt_2}+\left(\frac{1}{16}+
\frac{\bt_1+\bt_2}{12}+\frac{\bt_1^2+\bt_1\bt_2+\bt_2^2}{24}\right)\gs^2+\cO(\gs^4)\right].
\end{aligned} 
\label{eq:JT-12}
\end{equation}
These expressions are nothing but
\eqref{eq:WK-12} evaluated at the special values \eqref{eq:u0I_JT}.
One can check that the loop equation \eqref{eq:loopgen} for $n=0$
is satisfied with \eqref{eq:VeffJT} and \eqref{eq:JT-12},
though this is evident from the previous example with general $t_k$.

Another interesting example is what is called
the Airy case corresponding to
the trivial background $t_k=0~(k\geq0)$. 
In this case $u_0=I_n=0$ and $V_{\text{eff}}'(\xi)$
in \eqref{eq:dVeff} becomes
\begin{equation}
\begin{aligned}
 V_{\text{eff}}'(\xi)=2\rt{2\xi}.
\end{aligned} 
\label{eq:V-ary}
\end{equation}
The $n$-boundary correlators in the Airy case for $n=1,2,3$ are known in a closed form
\cite{okounkov2002generating,Beccaria:2020ykg,Okuyama:2021cub}
\begin{equation}
\begin{aligned}
 Z_1(\bt)&=\frac{1}{\rt{2\pi\bt^3}\gs}e^{\frac{\gs^2\bt^3}{24}},\\
Z_2(\bt_1,\bt_2)&=Z_1(\bt_1+\bt_2)\text{Erf}\left(\frac{\gs}{2\rt{2}}
\rt{\bt_1\bt_2(\bt_1+\bt_2)}\right),\\
Z_3(\bt_1,\bt_2,\bt_3)&=Z_1\Biggl(\sum_{i=1}^3\bt_i\Biggr)\\
&\hspace{1em}\times\Biggl[1-
4T\left(\frac{\gs}{2}\rt{\bt_1(\bt_2+\bt_3)(\bt_1+\bt_2+\bt_3)},\rt{\frac{\bt_2\bt_3}{\bt_1(\bt_1+\bt_2+\bt_3)}}\right)\\
&\hspace{3.31em}-4T\left(\frac{\gs}{2}\rt{\bt_2(\bt_3+\bt_1)(\bt_1+\bt_2+\bt_3)},\rt{\frac{\bt_3\bt_1}{\bt_2(\bt_1+\bt_2+\bt_3)}}\right)\\
&\hspace{3.31em}-4T\left(\frac{\gs}{2}\rt{\bt_3(\bt_1+\bt_2)(\bt_1+\bt_2+\bt_3)},\rt{\frac{\bt_1\bt_2}{\bt_3(\bt_1+\bt_2+\bt_3)}}\right)\Biggr],
\end{aligned} 
\label{eq:airy-123}
\end{equation}
where $\text{Erf}(z)$ and $T(z,a)$ denote the error function
and the Owen's $T$-function respectively
\begin{equation}
\begin{aligned}
 \text{Erf}(z)=\frac{2}{\rt{\pi}}\int_0^z dt \,e^{-t^2},\quad 
T(z,a)=\frac{1}{2\pi}\int_0^a dt\frac{e^{-\hf z^2(1+t^2)}}{1+t^2}.
\end{aligned} 
\end{equation} 
We have checked that the loop equations \eqref{eq:loopgen}
for $n=0,1$ are indeed satisfied by the 
correlators in the Airy case \eqref{eq:airy-123} with $V_{\text{eff}}'(\xi)$
in \eqref{eq:V-ary}.

%%%%%%%%%%%%%%%%%%%%%%%%%%%%%%%%%%%%%%%%%%%%%%%%%%%%%%%%%%%%%%%%%%%%%%%%
\section{Relation to other approaches}\label{sec:relation}
%%%%%%%%%%%%%%%%%%%%%%%%%%%%%%%%%%%%%%%%%%%%%%%%%%%%%%%%%%%%%%%%%%%%%%%%

\subsection{Free boson/fermion representation}\label{sec:freerep}

It is well-known that
$\tau=e^F$ of Witten-Kontsevich topological gravity is the tau-function
of the KdV hierarchy and it
has a free boson/fermion representation
(see e.g.~\cite{BBT,Aganagic:2003qj,Kostov:2009nj,Kostov:2010nw} and references therein)
\begin{equation}
\begin{aligned}
e^F=\bra t|V\ket,
\end{aligned} 
\label{eq:tau-WK}
\end{equation}
where the state $\bra t|$ is given by the coherent state of free boson
\begin{equation}
\begin{aligned}
 \bra t|=\bra \Om|\exp\left(\sum_{k=0}^\infty \frac{t_k\al_{2k+1}}{\gs(2k+1)!!}\right)
\end{aligned} 
\label{eq:bra-t}
\end{equation}
with $\al_n$ obeying the usual commutation relation of the free boson
\begin{equation}
\begin{aligned}
 {[}\al_n,\al_m{]}=n\cob_{n+m,0},\qquad \bra \Om|\al_n=0\quad (n<0).
\end{aligned} 
\label{eq:al-com}
\end{equation}
Note that only the odd modes $\al_{2k+1}$ of $\al_n$ appear
in \eqref{eq:bra-t} since the KdV hierarchy is a mod-2 reduction
of the KP hierarchy.

The state $|V\ket$ in \eqref{eq:tau-WK}
is written in terms of the free fermions $\psi_r,\psi_r^*~(r\in\mathbb{Z}+\hf)$
obeying the anti-commutation relation
\begin{equation}
\begin{aligned}
 \{\psi_r,\psi_s^*\}=\cob_{r+s,0},\qquad
 \psi_r|\Om\ket=\psi^*_r|\Om\ket=0\quad (r>0).
\end{aligned} 
\end{equation}
They are related to $\al_n$ by the usual bosonization
\begin{equation}
\begin{aligned}
 \al_n=\sum_{r\in\mathbb{Z}+\hf}:\psi_r\psi_{n-r}^*:.
\end{aligned} 
\end{equation}
Then $|V\ket$ is written as 
\begin{equation}
\begin{aligned}
 |V\ket=\exp\left(\sum_{m,n=0}^\infty A_{m,n}\psi_{-m-\hf}\psi^*_{-n-\hf}\right)|\Om\ket.
\end{aligned} 
\label{eq:V-ket}
\end{equation}
The generating function of $A_{m,n}$ for the Witten-Kontsevich $\tau$-function
is obtained in \cite{zhou2013explicit,zhou2015emergent,balogh2017geometric}:
\begin{equation}
\begin{aligned}
 \sum_{m,n=0}^\infty A_{m,n}z^{-m-1}w^{-n-1}=\frac{1}{z-w}+\frac{a(w)b(-z)-a(-z)b(w)}{z^2-w^2},
\end{aligned} 
\label{eq:A-gen}
\end{equation}
where $a(z)$ and $b(z)$ are given by
\begin{equation}
\begin{aligned}
 a(z)&=\sum_{m=0}^\infty 
\left(\frac{-\gs}{288}\right)^{m}\frac{(6m)!}{(2m)!(3m)!}z^{-3m},\\
b(z)&=-\sum_{m=0}^\infty \left(\frac{-\gs}{288}\right)^{m}\frac{(6m)!}{(2m)!(3m)!}
\frac{6m+1}{6m-1}z^{-3m+1}.
\end{aligned} 
\label{eq:azbz-def}
\end{equation}

Note that the derivative $\del_k$ with respect to the coupling $t_k$
is mapped to the operator $\al_{2k+1}$ when acting on the state $\bra t|$
in \eqref{eq:bra-t}
\begin{equation}
\begin{aligned}
 \del_k\bra t|=\frac{1}{\gs(2k+1)!!}\bra t|\al_{2k+1},\quad(k\geq0).
\end{aligned}
\label{eq:t-map} 
\end{equation}
Using the commutation relation \eqref{eq:al-com} one can also show that
\begin{equation}
\begin{aligned}
t_k\bra t|&=\gs(2k-1)!!\bra t|\al_{-2k-1}\\
&=\gs\frac{(-1)^k}{(-2k-1)!!}\bra t|\al_{-2k-1}
,\quad(k\geq0),
\end{aligned} 
\label{eq:t-map2} 
\end{equation}
where we have used $(2k-1)!!(-2k-1)!!=(-1)^k$.
Then the Virasoro constraint \eqref{eq:Vircond} with $L_m$ in \eqref{eq:Lm}
can be translated to the free boson/fermion language
via the dictionary \eqref{eq:t-map}, \eqref{eq:t-map2}
\begin{equation}
\begin{aligned}
 \h{L}_n|V\ket=0\quad(n\geq-1),
\end{aligned} 
\label{eq:Vir-con}
\end{equation}
where the Virasoro generator $\h{L}_n$ is given by
\begin{equation}
\begin{aligned}
\h{L}_n&=\qu\sum_{k\in\mathbb{Z}}:\al_{2k+1}\al_{2n-2k-1}:
+\frac{1}{16}\cob_{n,0}-\frac{1}{2\gs}\al_{2n+3}.
\end{aligned} 
\label{eq:Ln-def}
\end{equation}
Note that the linear term $-\frac{1}{2\gs}\al_{2n+3}$ in \eqref{eq:Ln-def}
arises from the shift of 
$\tilt_1=t_1-1$ \cite{Kac:1991nv,Itzykson:1992ya}.
Another useful expression of $|V\ket$ is the cut-and-join representation found in
\cite{Alexandrov:2010bn}
\begin{equation}
\begin{aligned}
 |V\ket&=e^W|\Om\ket,\\
W&=\frac{2\gs}{3}\sum_{m=-1}^\infty\al_{-2m-3}\h{L}_m',
\end{aligned}
\label{eq:V-candj} 
\end{equation}
where $\h{L}_m'=\h{L}_m+\frac{1}{2\gs}\al_{2m+3}$.
See appendix \ref{sec:V} for a derivation of this expression.

The boundary creation operator $\cB(\bt)$ in \eqref{eq:hBcB}
can also be translated to the free boson/fermion language as
\begin{equation}
\begin{aligned}
 \cB(\bt)\bra t|&=\bra t|\h{Z}(\bt),
\end{aligned} 
\end{equation}
where $\h{Z}(\bt)$ is given by
\begin{equation}
\begin{aligned}
 \h{Z}(\bt)&=\frac{1}{\rt{2\pi}}\sum_{k=-\infty}^\infty 
\frac{\bt^{k+\hf}}{(2k+1)!!}\til{\al}_{2k+1}.
\end{aligned} 
\end{equation}
Here $\til{\al}_n$ is defined by
\begin{equation}
\begin{aligned}
 \til{\al}_n=\al_n-\gs^{-1}\cob_{n,-3}=e^{-\frac{1}{3\gs}\al_3}\al_ne^{\frac{1}{3\gs}\al_3},
\end{aligned} 
\end{equation}
which is related to the shift of $\tilt_k=t_k-\cob_{k,1}$.
One can show that the operators $\h{Z}(\bt)$ mutually commute \cite{Okuyama:2020ncd}
\begin{equation}
\begin{aligned}
 {[}\h{Z}(\bt),\h{Z}(\bt'){]}=0,
\end{aligned} 
\end{equation}
and the full correlator \eqref{eq:genZn} is written as
\begin{equation}
\begin{aligned}
 \bra Z(\bt_1)\cdots Z(\bt_n)\ket=\frac{\bra t|\h{Z}(\bt_1)\cdots\h{Z}(\bt_n)|V\ket}{\bra t|V\ket}.
\end{aligned} 
\label{eq:corr-hatZ}
\end{equation}

In our previous paper \cite{Okuyama:2020ncd}, we identified the state $|V\ket$
as the Hartle-Hawking state $|\text{HH}\ket$ \cite{Hartle:1983ai}
\begin{equation}
\begin{aligned}
 |\text{HH}\ket=|V\ket.
\end{aligned} 
\label{eq:our-HH}
\end{equation}
This is based on the argument in \cite{Polchinski:1989fn}
that the Hartle-Hawking state 
is ``the most symmetric state.'' Indeed, $|V\ket$ is invariant
under the Virasoro generators \eqref{eq:Ln-def} and
$|V\ket$ can be thought of as the $\grp{SL}(2,\mathbb{R})$
invariant vacuum.
In particular, the constraint $\h{L}_0|V\ket=0$ corresponds to the Wheeler-DeWitt
equation.
This indicates that the state $|V\ket$ is a natural candidate
for the Hartle-Hawking state $|\text{HH}\ket$.

Our \eqref{eq:our-HH} is 
consistent with the identification
of the one-point function $\bra Z(\bt)\ket$
as the wavefunction of the Hartle-Hawking state, which is commonly adopted 
in 2d gravity literature
(see e.g.~\cite{Ginsparg:1993is} for a review)
\begin{equation}
\begin{aligned}
 \bra Z(\bt)\ket=\Psi_{\text{HH}}(\bt)=\bra Z(\bt)|\text{HH}\ket,
\end{aligned} 
\end{equation}
where $\bra Z(\bt)|$ is given by
\begin{equation}
\begin{aligned}
 \bra Z(\bt)|=\frac{\bra t|\h{Z}(\bt)}{\bra t|\text{HH}\ket}.
\end{aligned} 
\end{equation}
More generally, the multi-point correlator \eqref{eq:corr-hatZ} is written as
\begin{equation}
\begin{aligned}
\bra Z(\bt_1)\cdots Z(\bt_n)\ket&=\bra Z(\bt_1)\cdots Z(\bt_n)|\text{HH}\ket,
\end{aligned} 
\label{eq:HH-corr}
\end{equation}
where $\bra Z(\bt_1)\cdots Z(\bt_n)|$ is given by
\begin{equation}
\begin{aligned}
\bra Z(\bt_1)\cdots Z(\bt_n)|&=\frac{\bra t|\h{Z}(\bt_1)\cdots \h{Z}(\bt_n)}{\bra t|\text{HH}\ket}.
\end{aligned} 
\end{equation}

In \cite{Sen:1990rz,Imbimbo:1990ua},
the Virasoro constraint of matrix model
is interpreted as the gauge symmetry of
closed string field theory in a minimal model background. 
This suggests that the Virasoro constraint 
is the analogue of the bulk diffeomorphism invariance.
Since the loop equation \eqref{eq:loopgen} is equivalent to the Virasoro constraint,
one can regard the loop equation \eqref{eq:loopgen}
as a manifestation of the bulk diffeomorphism invariance.
The loop equation \eqref{eq:loopgen} relates the amplitudes
with different number of boundaries. This can be thought of as a 
gauge redundancy due to the ``large'' diffeomorphism relating different topologies
of spacetime \cite{Jafferis:2017tiu}.
In the language of \cite{Marolf:2020xie}, the Virasoro constraint
defines a null state
\begin{equation}
\begin{aligned}
 |\cN\ket&=\h{L}(\bt)|\text{HH}\ket=0,
\end{aligned} 
\label{eq:null}
\end{equation}
where $\h{L}(\bt)$ is obtained from \eqref{eq:Lofb} as
\begin{equation}
\begin{aligned}
 \h{L}(\bt)=\sum_{m=-1}^\infty\frac{\bt^{m+1}}{(m+1)!2^m}\h{L}_m.
\end{aligned} 
\end{equation}
More generally, acting $\h{Z}(\bt)$'s on $|\cN\ket$ 
also gives rise to a null state
\begin{equation}
\begin{aligned}
 \h{Z}(\bt_1)\cdots \h{Z}(\bt_n)|\cN\ket=
\h{Z}(\bt_1)\cdots \h{Z}(\bt_n)\h{L}(\bt)|\text{HH}\ket=0.
\end{aligned} 
\label{eq:null-n}
\end{equation}
As we have seen in the previous section, after rewriting \eqref{eq:null-n}
as
\begin{equation}
\begin{aligned}
 \h{L}(\bt)\h{Z}(\bt_1)\cdots \h{Z}(\bt_n)|\text{HH}\ket-\sum_{i=1}^n
\h{Z}(\bt_1)\cdots[\h{L}(\bt),\h{Z}(\bt_i)]\cdots \h{Z}(\bt_n)|\text{HH}\ket=0,
\end{aligned} 
\end{equation}
and using $[\h{L}(\bt),\h{Z}(\bt')]=-\bt'\h{Z}(\bt+\bt')$,
\eqref{eq:null-n} becomes equivalent to the loop equation \eqref{eq:loopgen}.
Thus we can regard the loop equation \eqref{eq:loopgen}
as the equation for the null state due to the large diffeomorphism invariance.

\subsection{Relation to Marolf-Maxfield \cite{Marolf:2020xie}}

Let us discuss the relation between 
our expression \eqref{eq:HH-corr} and the one proposed by Marolf and Maxfield
in \cite{Marolf:2020xie}
\begin{equation}
\begin{aligned}
 \bra Z(\bt_1)\cdots Z(\bt_n)\ket=
\frac{\bra\MMHH|\h{Z}(\bt_1)\cdots\h{Z}(\bt_n)|\MMHH\ket}{\bra\MMHH|\MMHH\ket}.
\end{aligned}
\label{eq:MM-prop}
\end{equation}
This is different from our \eqref{eq:HH-corr}.
In their formulation the Hartle-Hawking state
is represented by both the bra $\bra\MMHH|$ and ket $|\MMHH\ket$
as in \eqref{eq:MM-prop},
while in our formulation the bra and ket are treated asymmetrically
and the Hartle-Hawking state is represented by
the ket $|\text{HH}\ket$ only.
In other words, our expression \eqref{eq:HH-corr} corresponds to a special
(Euclidean) time-slicing of the spacetime where the initial state has no boundary 
and all the boundaries are on the final state. 
On the other hand, the proposal \eqref{eq:MM-prop} in \cite{Marolf:2020xie}
is based on a certain 
assumption of the cutting and sewing of the gravitational path integral
and the existence of the CPT conjugation.

However, as emphasized in \cite{Moore:1991ag}, 
the sewing of path integral in quantum gravity is 
quite different from the ordinary quantum field theories without gravity.
Let us recall the argument in \cite{Moore:1991ag}.
When the manifold $M$ is cut into two pieces $M_1$ and $M_2$,
the path integral of quantum fields
over $M$ is obtained by gluing $M_1$ and $M_2$ 
along the common boundary $\Si=\del M_1=\del M_2$
\begin{equation}
\begin{aligned}
 \int_M \cD\phi \,e^{-S(\phi)}=
\int_\Si \cD\phi_b \psi_{M_1}(\phi_b) \psi_{M_2}(\phi_b)=\bra\psi_{M_1}|\psi_{M_2}\ket,
\end{aligned} 
\label{eq:gluing}
\end{equation}
where $\psi_{M_i}(\phi_b)~(i=1,2)$ is the wavefunction defined by
the path integral over $M_i$ with the fixed boundary value $\phi|_\Si=\phi_b$  
\begin{equation}
\begin{aligned}
 \psi_{M_i}(\phi_b)=\int_{M_i;\phi|_\Si=\phi_b} \cD\phi \,e^{-S(\phi)}.
\end{aligned} 
\end{equation}
In the case of quantum gravity we have to perform the path integral
over the metrics, which in particular includes the integral over the 
moduli space of metrics.
Let $\cM$, $\cM_1$ and $\cM_2$
denote the moduli spaces of $M$, $M_1$ and $M_2$ respectively.
In calculating $\psi_{M_1}$ and $\psi_{M_2}$ we integrate over $\cM_1$ and $\cM_2$,
and as a consequence the inner product $\bra\psi_{M_1}|\psi_{M_2}\ket$
is given by the integral over $\cM_1\times\cM_2$.
However, $\cM$ is not equal to
the product of $\cM_1$ and $\cM_2$ in general 
\begin{equation}
\begin{aligned}
 \cM\ne \cM_1\times\cM_2.
\end{aligned} 
\end{equation}
Therefore, the inner product $\bra\psi_{M_1}|\psi_{M_2}\ket$ does not correspond to
the integral over $M$. In other words, the sewing operation
does not commute with the integration over the moduli \cite{Moore:1991ag}.
As mentioned in \cite{Moore:1991ag}, the sewing operation 
in the gravitational path integral
is valid only for a given point in moduli space
and it breaks down when we integrate over the moduli space.
The sewing operation works for local fields on spacetime,
but the moduli space is defined from the global property
of spacetime.

Of course, one can also consider the cutting and sewing of the moduli space integral.
For instance, the Weil-Petersson volume of the moduli space of Riemann surfaces
satisfies the recursion relation found by Mirzakhani 
\cite{mirzakhani2007simple},\footnote{
As shown by Eynard and Orantin \cite{Eynard:2007fi}, 
this recursion relation is equivalent to the topological recursion of the double-scaled matrix model.
}
which comes from the pant decomposition of the underlying Riemann surfaces.
This recursion relation essentially says that 
the higher genus Weil-Petersson volume is obtained by summing over
all possible pant decompositions.
In particular, we have to include the contribution of
a pair of pants
connecting two punctures on either connected or disconnected 
Riemann surfaces with lower genera, which can be thought of as a contribution of 
wormhole connecting either the same universe or disjoint universes.\footnote{
Here, by ``wormhole'' we mean a wormhole with a puncture, which
is topologically equivalent to a pair of pants.}
In other words, in 2d gravity
we have to sum over all possible cutting/sewing of the spacetime
and include the contribution of wormholes.
We expect that this is a general feature of the gravitational path integral.

The above argument urges us
to reconsider the derivation of
the expression \eqref{eq:MM-prop} in \cite{Marolf:2020xie} more carefully.\footnote{
The naive applications of sewing operation
in the third quantization of universes \cite{Strominger:1988ys}
or ``universe field theories'' (see \cite{Anous:2020lka,Casali:2021ewu}
and references therein)
suffer from the same problem.
}
In fact, the correlator in 2d quantum gravity, which is exactly solved in terms of the
double-scaled matrix model, takes the form \eqref{eq:HH-corr}, not
\eqref{eq:MM-prop}.
The Hilbert space
based on the conventional free boson/fermion representation
that we studied in section~\ref{sec:freerep}
does not seem to be identical with the one
proposed in \cite{Marolf:2020xie} and we
do not know how to relate our \eqref{eq:HH-corr} with \eqref{eq:MM-prop}
proposed in \cite{Marolf:2020xie}.
We leave this as an important future problem.

\subsection{Relation to Ishibashi-Kawai \cite{Ishibashi:1993nq,Ishibashi:1995np}}
Next we consider the relation to the closed string field theory (SFT) of
non-critical strings developed in a series of papers by
Ishibashi and Kawai \cite{Ishibashi:1993nq,Ishibashi:1995np}.
Their SFT naturally arises in the quantization of 2d gravity
in the temporal gauge \cite{Ikehara:1994xs}.
Let us briefly recall the formalism of \cite{Ishibashi:1993nq,Ishibashi:1995np}.
First they introduce the creation and annihilation operators
$\Psi^\dag(\bt),\Psi(\bt)$ of the macroscopic loops
obeying the commutation relation
\begin{equation}
\begin{aligned}
{}[\Psi(\bt),\Psi^\dag(\bt')]=\cob(\bt-\bt'),
\end{aligned} 
\end{equation}
and define the ``vacuum state'' $|0\ket$ as
\begin{equation}
\begin{aligned}
 \Psi(\bt)|0\ket=\bra0|\Psi^\dag(\bt)=0.
\end{aligned} 
\end{equation}
Then the partition function $e^F$ is written as
\begin{equation}
\begin{aligned}
 e^F=\bra 0|v\ket,
\end{aligned} 
\end{equation}
where $|v\ket$ is given by
\begin{equation}
\begin{aligned}
 |v\ket=\lim_{\tau\to\infty}e^{-\tau \cH}|0\ket.
\end{aligned} 
\label{eq:v-ket}
\end{equation}
Here $\cH$ can be thought of as 
the Fokker-Planck Hamiltonian for the stochastic quantization
\cite{Ikehara:1994xs}. $\cH$ describes the splitting and joining of loops
and its explicit form is given by
\begin{equation}
\begin{aligned}
 \cH=\int_0^\infty d\bt\bt\Psi^\dag(\bt)\left[T(\bt)+\bt^{-1}\rho(\bt)\right],
\end{aligned} 
\label{eq:H-SFT}
\end{equation}
with
\begin{equation}
\begin{aligned}
 T(\bt)=\int_0^\bt ds \Psi(s)\Psi(\bt-s)
+\gs^2\int_0^\infty ds\, s\Psi^\dag(s)\Psi(\bt+s).
\end{aligned} 
\end{equation}
This satisfies the continuum Virasoro algebra
\begin{equation}
\begin{aligned}
{}[T(\bt),T(\bt')]=\gs^2(\bt-\bt')T(\bt+\bt').
\end{aligned} 
\end{equation} 
$\rho(\bt)$ in \eqref{eq:H-SFT} represents the tadpole term.
In this formalism, the correlator of macroscopic loops
is written as
\begin{equation}
\begin{aligned}
 \bra Z(\bt_1)\cdots Z(\bt_n)\ket
=\gs^{-n}\frac{\bra 0|\Psi(\bt_1)\cdots\Psi(\bt_n)|v\ket}{\bra 0|v\ket}.
\end{aligned} 
\end{equation}

It is argued in \cite{Ishibashi:1993nq,Ishibashi:1995np} that the state $|v\ket$
in \eqref{eq:v-ket} satisfies the constraint
\begin{equation}
\begin{aligned}
 \left[T(\bt)+\bt^{-1}\rho(\bt)\right]|v\ket=0.
\end{aligned} 
\label{eq:v-const}
\end{equation}
This condition fixes the form of $\rho(\bt)$ in terms of
the disk amplitude $Z_1^{g=0}(\bt)$
\begin{equation}
\begin{aligned}
\bt^{-1}\rho(\bt)=-\gs^2\int_0^\bt dsZ_1^{g=0}(s)Z_1^{g=0}(\bt-s).
\end{aligned} 
\label{eq:rho-bt}
\end{equation}
Note that this is just a formal expression since this integral is divergent
due to the negative powers of $s$ in $Z_1^{g=0}(s)$ (see \eqref{eq:Z1g0ter}).
From \eqref{eq:v-const} the following form of the loop equation
is obtained in \cite{Ishibashi:1993nq,Ishibashi:1995np}
\begin{equation}
\begin{aligned}
 \int_0^\bt ds\bra\Psi(s)\Psi(\bt-s)\ket_J+\gs^2
\int_0^\infty ds sJ(s)\bra\Psi(\bt+s)\ket_J
+\bt^{-1}\rho(\bt)=0,
\end{aligned} 
\label{eq:loop-IK}
\end{equation}
where $\bra \cdots\ket_J$ is defined by
\begin{equation}
\begin{aligned}
 \bra\cdots\ket_J=\frac{\bra0|(\cdots)\exp\big[\int_0^\infty dsJ(s)\Psi(s)\big]|v\ket}{\bra0|\exp\big[\int_0^\infty dsJ(s)\Psi(s)\big]|v\ket}.
\end{aligned} 
\end{equation}

The loop equation \eqref{eq:loop-IK} is almost identical to
our result, but there are some subtle differences.
Let us compare \eqref{eq:loop-IK} and the free boson/fermion formalism.
One can formally introduce the ``boundary annihilation operator''
\begin{equation}
\begin{aligned}
 \h{Z}^\dag(\bt)&=\lim_{\ve\to +0}\frac{\ri}{\rt{2\pi}}
\sum_{k=0}^\infty(-1)^k\Biggl[(\bt+\ri\ve)^{k+\hf}\frac{\al_{2k+1}}{(2k+1)!!}
-(\bt-\ri\ve)^{-k-\hf}\frac{\al_{-2k-1}}{(-2k-1)!!}\Biggr],
\end{aligned} 
\end{equation}
which satisfies
\begin{equation}
\begin{aligned}
{}[\h{Z}(\bt),\h{Z}^\dag(\bt')]=\bt\cob(\bt-\bt').
\end{aligned} 
\end{equation}
Then it is tempting to identify
\begin{equation}
\begin{aligned}
 \Psi(\bt)&\leftrightarrow \gs\h{Z}(\bt),\\
\Psi^\dag(\bt) &\leftrightarrow \frac{1}{\gs\bt}\h{Z}^\dag(\bt).
\end{aligned} 
\end{equation}
This identification works at the level of commutation relation, but
the Hilbert spaces on which these operators act are different.
In particular, there is no such state $|0\ket$
annihilated by $\h{Z}(\bt)$ for all $\bt\geq0$
at least in the Fock space of free boson/fermion.

Also, it is argued in
\cite{Ishibashi:1995np} that the algebra of 
$\til{T}(\bt)=T(\bt)+\bt^{-1}\rho(\bt)$ does not close
due to the presence of the tadpole term $\rho(\bt)$
\begin{equation}
\begin{aligned}
{}[\til{T}(\bt_1),\til{T}(\bt_2)]=\gs^2(\bt_1-\bt_2)\til{T}(\bt_1+\bt_2)-
\gs^2\frac{\bt_1-\bt_2}{\bt_1+\bt_2}\rho(\bt_1+\bt_2),
\end{aligned} 
\end{equation}
and the consistency of the constraint $\til{T}(\bt)|v\ket=0$ in \eqref{eq:v-const}
is a subtle issue. 
On the other hand, in our case the constraint algebra is closed
\begin{equation}
\begin{aligned}
{}[\h{L}(\bt_1),\h{L}(\bt_2)]|V\ket=(\bt_1-\bt_2)\h{L}(\bt_1+\bt_2)|V\ket=0
\end{aligned} 
\end{equation}
and there is no problem associated with the tadpole term.
This difference can be traced back to the fact that
only the positive powers of $\bt$ appear in the definition of
$L(\bt)$ in \eqref{eq:Lofb} and the convolution of the disk amplitudes 
in \eqref{eq:rho-bt} is
already subtracted from the beginning (see \eqref{eq:cL2})
\begin{equation}
\begin{aligned}
 L(\bt)=\int_0^\bt ds B(s)B(\bt-s)+\cdots\ne
\int_0^\bt ds \cB(s)\cB(\bt-s)+\cdots.
\end{aligned} 
\end{equation}
Note that $B(\bt)$ denotes the universal part which contains only 
the positive powers of $\bt$.

It is interesting to observe that the cut-and-join operator $W$ in 
\eqref{eq:V-candj} has a similar form with the SFT Hamiltonian $\cH$
in \eqref{eq:H-SFT}. $W$ in 
\eqref{eq:V-candj} can be thought of as the SFT Hamiltonian written in terms
of the microscopic loop operators $\al_n$.

%%%%%%%%%%%%%%%%%%%%%%%%%%%%%%%%%%%%%%%%%%%%%%%%%%%%%%%%%%%%%%%%%%%%%%%%
\section{Conclusions and outlook}\label{sec:conclusion}
%%%%%%%%%%%%%%%%%%%%%%%%%%%%%%%%%%%%%%%%%%%%%%%%%%%%%%%%%%%%%%%%%%%%%%%%

In this paper we have presented a detailed proof of the loop equations
\eqref{eq:loopfund} and \eqref{eq:loopgen} obeyed by the multi-boundary
correlators in Witten-Kontsevich topological gravity
with arbitrary background $\{t_k\}$.
Since the Virasoro operator $L(\bt)$ in \eqref{eq:Lofb}
contains only nonnegative powers of $\bt$, 
the disk amplitude $Z_1^{g=0}(\bt)$ does not appear in the 
convolution part (i.e.~the $s$-integral) 
of \eqref{eq:loopfund} and \eqref{eq:loopgen}.
We emphasize that our loop equations
\eqref{eq:loopfund} and \eqref{eq:loopgen} 
are valid for the general background $\{t_k\}$ including the $u_0\ne0$ case,
which has not been worked out in the literature before.
As a concrete example, we have demonstrated that our loop equations are
indeed satisfied for JT gravity and the Airy case.

One of the motivations of our study of the loop equation is 
to understand the relation to the discussion of the null state
by Marolf and Maxfield \cite{Marolf:2020xie}.
In section \ref{sec:relation}, we have argued that 
our loop equation has an interpretation as the null state in the free boson/fermion
language of the Witten-Kontsevich $\tau$-function.
It is interesting that the loop equation relates
the multi-boundary correlators with different number of boundaries
and it can be thought of as a consequence of the ``large'' diffeomorphism 
relating different topologies of spacetime \cite{Jafferis:2017tiu}.

As we mentioned in section \ref{sec:relation}, our expression of multi-boundary correlator
\eqref{eq:HH-corr} is different from \eqref{eq:MM-prop}
of Marolf and Maxfield \cite{Marolf:2020xie}.
Since 2d gravity is completely solved by the double-scaled matrix model and
\eqref{eq:HH-corr} is what we get from the general formula
of Witten-Kontsevich topological gravity,
we have to take the result \eqref{eq:HH-corr} very seriously.
The integrable structure of Witten-Kontsevich topological gravity
leading to the result \eqref{eq:HH-corr} is tightly constrained and there
is no natural way to rewrite \eqref{eq:HH-corr} into the form of \eqref{eq:MM-prop}. 
We suspect that one 
of the possible pitfalls of the discussion in \cite{Marolf:2020xie} is the naive
application of the sewing operation in the gravitational path integral.
As discussed in \cite{Moore:1991ag}, the sewing operation breaks down when
we integrate over the moduli space of metrics.
It would be desirable to reconsider the argument in \cite{Marolf:2020xie}
in view of the remark in \cite{Moore:1991ag}.
We leave this as an important future problem.

\acknowledgments
This work was supported in part by JSPS KAKENHI Grant
Nos.~19K03845 and 19K03856,
and JSPS Japan-Russia Research Cooperative Program.

%%%%%%%%%%%%%%%%%%%%%%%%%%%%%%%%%%%%%%%%%%%%%%%%%%%%%%%%%%%%%%%%%%%%%%%%
%%%%%%%%%%%%%%%%%%%%%%%%%%%%%%%%%%%%%%%%%%%%%%%%%%%%%%%%%%%%%%%%%%%%%%%%
\appendix

%%%%%%%%%%%%%%%%%%%%%%%%%%%%%%%%%%%%%%%%%%%%%%%%%%%%%%%%%%%%%%%%%%%%%%%%
\section{Proof of commutation relation \eqref{eq:LBcomm}}
\label{sec:LB}
%%%%%%%%%%%%%%%%%%%%%%%%%%%%%%%%%%%%%%%%%%%%%%%%%%%%%%%%%%%%%%%%%%%%%%%%

In this section we prove the commutation relation \eqref{eq:LBcomm}.
The l.h.s.~of \eqref{eq:LBcomm} is written as
\begin{align}
\begin{aligned}
{}[\vL(\beta),\cB(\beta')]
&=\!\sum_{m=-1}^\infty\frac{\beta^{m+1}}{(m+1)!2^m}
\frac{1}{\sqrt{2\pi}}
 \sum_{k=0}^\infty\left(\gs^{-1}(-1)^k{\beta'}^{-k-\frac{1}{2}}
 [L_m,\tilt_k]
 +\gs{\beta'}^{k+\frac{1}{2}}[L_m,\partial_k]\right).
\end{aligned}
\end{align}
One can show that
\begin{align}
\begin{aligned}
{}[L_m,\tilt_k]
 &=\left\{\begin{array}{ll}
  \dfrac{1}{2}\dfrac{(2k+1)!!}{(2k-2m-1)!!}\tilt_{k-m}&(m\le k),\\[2ex]
  \dfrac{\gs^2}{2}(2k+1)!!(2m-2k-1)!!\partial_{m-k-1}\quad&(m\ge k+1),
	  \end{array}\right.\\
{}[L_m,\partial_k]
 &=\left\{\begin{array}{ll}
  -\dfrac{t_0}{2\gs^2}&(m,k)=(-1,0),\\[2ex]
  -\dfrac{1}{2}\dfrac{(2k+2m+1)!!}{(2k-1)!!}\partial_{k+m}\quad
  &\mbox{otherwise}.
	  \end{array}\right.
\end{aligned}
\end{align}
Using these relations we obtain
\begin{align}
\begin{aligned}
&[\vL(\beta),\cB(\beta')]\\
&=\frac{1}{\sqrt{2\pi}\gs}
 \sum_{m=-1}^\infty\sum_{k=m}^\infty
 \frac{\beta^{m+1}(-1)^k{\beta'}^{-k-\frac{1}{2}}}{(m+1)!2^{m+1}}
 \frac{(2k+1)!!}{(2k-2m-1)!!}\tilt_{k-m}\\
&\hspace{1em}+\frac{\gs}{\sqrt{2\pi}}
 \sum_{m=1}^\infty\sum_{k=0}^{m-1}
 \frac{\beta^{m+1}(-1)^k{\beta'}^{-k-\frac{1}{2}}}{(m+1)!2^{m+1}}
 (2k+1)!!(2m-2k-1)!!\partial_{m-k-1}\\
&\hspace{1em}-\frac{\gs}{\sqrt{2\pi}}
 \sum_{\substack{m\ge-1,\ k\ge 0\\[1ex] (m,k)\ne (-1,0)}}
 \frac{\beta^{m+1}{\beta'}^{k+\frac{1}{2}}}{(m+1)!2^{m+1}}
 \frac{(2k+2m+1)!!}{(2k-1)!!}\partial_{k+m}.
\end{aligned}
\label{eq:LBcommbis}
\end{align}
Note that the contribution of the $(m,k)=(-1,0)$ case of
$[L_m,\partial_k]$ is included as
the $(m,k)=(-1,-1)$ case of the first term.

By setting $\tm=m+1$ and $\ell=k-m$
the first term of \eqref{eq:LBcomm} is rewritten as
\begin{align}
\begin{aligned}
&\frac{1}{\sqrt{2\pi}\gs}
 \sum_{m=-1}^\infty\sum_{k=m}^\infty
 \frac{\beta^{m+1}(-1)^k{\beta'}^{-k-\frac{1}{2}}}{(m+1)!2^{m+1}}
 \frac{(2k+1)!!}{(2k-2m-1)!!}\tilt_{k-m}\\
&=-\beta'\frac{1}{\sqrt{2\pi}\gs}
 \sum_{\ell=0}^\infty(-1)^\ell{\beta'}^{-\ell-\frac{1}{2}}\tilt_\ell
 \sum_{\tm=0}^\infty\frac{1}{\tm}\left(\frac{\beta}{\beta'}\right)^\tm
 \frac{(-1)^\tm(2\ell+\tm-1)!!}{2^\tm (2\ell-1)!!}\\
&=-\beta'\frac{1}{\sqrt{2\pi}\gs}
 \sum_{\ell=0}^\infty(-1)^\ell{\beta'}^{-\ell-\frac{1}{2}}\tilt_\ell
 \left(1+\frac{\beta}{\beta'}\right)^{-\ell-\frac{1}{2}}
 =-\beta'\cB(\beta+\beta')\Big|_{\tilt}.
\end{aligned}
\label{eq:LBcommbis1}
\end{align}
This reproduces the polynomial part of the r.h.s.~of \eqref{eq:LBcomm}.

Next, by setting $\ell=m-k-1$
the second term of \eqref{eq:LBcommbis} becomes
\begin{align}
\begin{aligned}
&\frac{\gs}{\sqrt{2\pi}}
 \sum_{m=1}^\infty\sum_{k=0}^{m-1}
 \frac{\beta^{m+1}(-1)^k{\beta'}^{-k-\frac{1}{2}}}{(m+1)!2^{m+1}}
 (2k+1)!!(2m-2k-1)!!\partial_{m-k-1}\\
&=\frac{\gs}{\sqrt{2\pi}}
 \sum_{m=1}^\infty\sum_{\ell=0}^{m-1}
 \frac{\beta^{m+1}(-1)^{m-\ell-1}{\beta'}^{\ell-m+\frac{1}{2}}}
      {(m+1)!2^{m+1}}
 (2m-2\ell-1)!!(2\ell+1)!!\partial_\ell\\
&=-\frac{\gs}{\sqrt{2\pi}}
 \sum_{\ell=0}^\infty\sum_{m=\ell+1}^\infty
 \frac{\beta^{m+1}(-1)^{m-\ell}{\beta'}^{\ell-m+\frac{1}{2}}}
      {(m+1)!2^{m+1}}
 (2m-2\ell-1)!!(2\ell+1)!!\partial_\ell\\
&=-\frac{\gs}{\sqrt{2\pi}}
 \sum_{\ell=0}^\infty\sum_{m=\ell+1}^\infty
 \frac{\beta^{m+1}{\beta'}^{\ell-m+\frac{1}{2}}}
      {(m+1)!}
 \frac{\Gamma(\ell+\frac{3}{2})}
      {\Gamma(\ell-m+\frac{1}{2})}\partial_\ell.
\end{aligned}
\label{eq:LBcommbis2}
\end{align}

On the other hand, by setting $\ell=k+m$
the third term of \eqref{eq:LBcommbis} becomes
\begin{align}
\begin{aligned}
&-\frac{\gs}{\sqrt{2\pi}}
 \sum_{\substack{m\ge-1,\ k\ge 0\\[1ex] (m,k)\ne (-1,0)}}
 \frac{\beta^{m+1}{\beta'}^{k+\frac{1}{2}}}{(m+1)!2^{m+1}}
 \frac{(2k+2m+1)!!}{(2k-1)!!}\partial_{k+m}\\
&=-\frac{\gs}{\sqrt{2\pi}}
 \sum_{\ell=0}^\infty\sum_{m=-1}^{\ell}
 \frac{\beta^{m+1}{\beta'}^{\ell-m+\frac{1}{2}}}
      {(m+1)!2^{m+1}}
 \frac{(2\ell+1)!!}{(2\ell-2m-1)!!}\partial_{\ell}\\
&=-\frac{\gs}{\sqrt{2\pi}}
 \sum_{\ell=0}^\infty\sum_{m=-1}^{\ell}
 \frac{\beta^{m+1}{\beta'}^{\ell-m+\frac{1}{2}}}
      {(m+1)!}
 \frac{\Gamma(\ell+\frac{3}{2})}
      {\Gamma(\ell-m+\frac{1}{2})}\partial_{\ell}.
\end{aligned}
\label{eq:LBcommbis3}
\end{align}
From \eqref{eq:LBcommbis2} and \eqref{eq:LBcommbis3}
we see that the sum of the second and the third terms
in \eqref{eq:LBcommbis} becomes (we set $\tm=m+1$)
\begin{align}
\begin{aligned}
&-\frac{\gs}{\sqrt{2\pi}}
 \sum_{\ell=0}^\infty\sum_{m=-1}^\infty
 \frac{\beta^{m+1}{\beta'}^{\ell-m+\frac{1}{2}}}
      {(m+1)!}
 \frac{\Gamma(\ell+\frac{3}{2})}
      {\Gamma(\ell-m+\frac{1}{2})}\partial_{\ell}\\
&=-\beta'\frac{\gs}{\sqrt{2\pi}}
 \sum_{\ell=0}^\infty{\beta'}^{\ell+\frac{1}{2}}
 \sum_{\tm=0}^\infty\frac{1}{\tm!}
 \left(\frac{\beta}{\beta'}\right)^\tm
 \frac{\Gamma(\ell+\frac{3}{2})}
      {\Gamma(\ell-\tm+\frac{3}{2})}\partial_{\ell}\\
&=-\beta'\frac{\gs}{\sqrt{2\pi}}
 \sum_{\ell=0}^\infty{\beta'}^{\ell+\frac{1}{2}}
 \left(1+\frac{\beta}{\beta'}\right)^{\ell+\frac{1}{2}}
 \partial_{\ell}\\
&=-\beta'\cB(\beta+\beta')\Big|_{\partial}.
\end{aligned}
\end{align}
This reproduces the derivative part of the r.h.s.~of \eqref{eq:LBcomm}.
Thus we have proved \eqref{eq:LBcomm}.

%%%%%%%%%%%%%%%%%%%%%%%%%%%%%%%%%%%%%%%%%%%%%%%%%%%%%%%%%%%%%%%%%%%%%%%%
\section{Proof of relation \eqref{eq:cLtdel}}\label{sec:prooftdel}
%%%%%%%%%%%%%%%%%%%%%%%%%%%%%%%%%%%%%%%%%%%%%%%%%%%%%%%%%%%%%%%%%%%%%%%%

In this section we prove the relation \eqref{eq:cLtdel}.
Let us start from the r.h.s.~of \eqref{eq:cLtdel}.
To evaluate the first term, we first substitute \eqref{eq:I0exp}
and $e^{\beta v}=\sum_{m=0}^\infty\frac{\beta^mv^m}{m!}$
into \eqref{eq:Z1univg0}. This gives
\begin{align}
\begin{aligned}
Z_1^{\univ,g=0}(\beta)
 &=\frac{1}{\gs}\sqrt{\frac{\beta}{2\pi}}
  \sum_{l=1}^\infty\frac{I_l-\delta_{l,1}}{l!}
  \sum_{m=0}^\infty\frac{\beta^m}{m!}
  \int_0^{u_0}dv (v-u_0)^l v^m\\
 &=\frac{1}{\sqrt{2\pi}\gs}
  \sum_{l=1}^\infty
  \sum_{m=0}^\infty\frac{(-1)^l (I_l-\delta_{l,1})}{(l+m+1)!}
  u_0^{l+m+1}\beta^{m+\frac{1}{2}}.
\end{aligned}
\end{align}
Using this expression and the definition \eqref{eq:Bop} of $B(\beta)$ 
we obtain
\begin{align}
\begin{aligned}
&2\int_0^\beta ds Z_1^{\univ,g=0}(\beta-s) B(s)\\
&=\frac{1}{\pi}
  \sum_{l=1}^\infty\sum_{m=0}^\infty\sum_{n=0}^\infty
  \frac{(-1)^l (I_l-\delta_{l,1})}{(l+m+1)!}
  u_0^{l+m+1}\int_0^\beta ds
  (\beta-s)^{m+\frac{1}{2}}s^{n+\frac{1}{2}}\partial_n.
\end{aligned}
\label{eq:Vircond1_3}
\end{align}
The last integral is evaluated as
\begin{align}
\begin{aligned}
\int_0^\beta ds(\beta-s)^{m+\frac{1}{2}}s^{n+\frac{1}{2}}
&=\frac{\Gamma(m+\frac{3}{2})\Gamma(n+\frac{3}{2})}{\Gamma(m+n+3)}
  \beta^{m+n+2}\\
&=\frac{(-1)^{m+1}\pi\Gamma(n+\frac{3}{2})}
       {\Gamma(-m-\frac{1}{2})(m+n+2)!}\beta^{m+n+2}.
\end{aligned}
\end{align}
Thus we have
\begin{align}
\begin{aligned}
&2\int_0^\beta ds Z_1^{\univ,g=0}(\beta-s) B(s)\\
&=\sum_{l=1}^\infty\sum_{m=0}^\infty\sum_{n=0}^\infty
  \frac{(-u_0)^{l+m+1}}{(l+m+1)!}
  \frac{I_l-\delta_{l,1}}{\Gamma(-m-\frac{1}{2})}
  \frac{\Gamma(n+\frac{3}{2})}{(m+n+2)!}
  \beta^{m+n+2}\partial_n\\
&=\sum_{n=0}^\infty\sum_{\tm=2}^\infty\sum_{l=1}^{\tm-1}
  \frac{(-u_0)^\tm}{\tm!}
  \frac{I_l-\delta_{l,1}}{\Gamma(l-\tm+\frac{1}{2})}
  \frac{\Gamma(n+\frac{3}{2})}{(n-l+\tm+1)!}
  \beta^{n-l+\tm+1}\partial_n.
\end{aligned}
\label{eq:Vircond1_3bis}
\end{align}
In the last step we have set $\tm=l+m+1$.

On the other hand, the second term on the r.h.s.~of \eqref{eq:cLtdel}
is evaluated as
\begin{align}
\begin{aligned}
&-\frac{1}{\gs}V'_{\rm eff}(\partial_\beta)B(\beta)\\
&=\sum_{l=1}^\infty\sum_{n=0}^\infty
  \frac{I_l-\delta_{l,1}}{\Gamma(l+\frac{1}{2})}
  (\partial_\beta-u_0)^{l-\frac{1}{2}}\beta^{n+\frac{1}{2}}
  \partial_n\\
&=\sum_{l=1}^\infty\sum_{n=0}^\infty
  \frac{I_l-\delta_{l,1}}{\Gamma(l+\frac{1}{2})}
  \sum_{m=0}^\infty
  \frac{\Gamma(l+\frac{1}{2})}{\Gamma(l-m+\frac{1}{2})}
  \frac{(-u_0)^m}{m!}\partial_\beta^{l-\frac{1}{2}-m}
  \beta^{n+\frac{1}{2}}
  \partial_n\\
&=\sum_{l=1}^\infty\sum_{n=0}^\infty
  \sum_{m=0}^\infty
  \frac{I_l-\delta_{l,1}}{\Gamma(l-m+\frac{1}{2})}
  \frac{(-u_0)^m}{m!}
  \frac{\Gamma(n+\frac{3}{2})}{\Gamma(n-l+m+2)}\beta^{n-l+m+1}
  \partial_n.
\end{aligned}
\label{eq:VeffZ1}
\end{align}
In the last step we have 
used the formula \eqref{eq:hfintdiff0}.

Now, observe that the summand of \eqref{eq:Vircond1_3bis}
is identical with that of \eqref{eq:VeffZ1}.
Therefore, subtracting \eqref{eq:Vircond1_3bis}
from \eqref{eq:VeffZ1} we see that
the r.h.s.~of \eqref{eq:cLtdel} becomes
\begin{align}
\begin{aligned}
&-2\int_0^\beta dsZ_1^{\univ,g=0}(\beta-s)B(s)
 -\frac{1}{\gs}V'_{\rm eff}(\partial_\beta)B(\beta)\\
&=\sum_{n=0}^\infty
  \sum_{\substack{m\ge 0,\ l\ge m\\[.5ex] (m,l)\ne (0,0)}}
  \frac{I_l-\delta_{l,1}}{\Gamma(l-m+\frac{1}{2})}
  \frac{(-u_0)^m}{m!}
  \frac{\Gamma(n+\frac{3}{2})}{\Gamma(n-l+m+2)}\beta^{n-l+m+1}
  \partial_n.
\end{aligned}
\label{eq:cLtdrhs1}
\end{align}
By adding and subtracting the $(m,l)=(0,0)$ contribution
and setting $k=l-m$,
\eqref{eq:cLtdrhs1} is rewritten as
\begin{align}
  \sum_{n=0}^\infty\left(\sum_{m=0}^\infty\sum_{k=0}^{n+1}
  \frac{I_{k+m}-\delta_{k+m,1}}{\Gamma(k+\frac{1}{2})}
  \frac{(-u_0)^m}{m!}
  \frac{\beta^{n-k+1}}{(n-k+1)!}
  -\frac{I_0\beta^{n+1}}{\Gamma(\frac{1}{2})(n+1)!}
  \right)
  \Gamma(n+\tfrac{3}{2})\partial_n.
\end{align}
By using \eqref{eq:tinI} this is rewritten as
\begin{align}
\begin{aligned}
 &\sum_{n=0}^\infty\left(\sum_{k=0}^{n+1}
  \frac{t_k-\delta_{k,1}+u_0\delta_{k,0}}{\Gamma(k+\frac{1}{2})}
  \frac{\beta^{n-k+1}}{(n-k+1)!}
  -\frac{I_0\beta^{n+1}}{\Gamma(\frac{1}{2})(n+1)!}
  \right)
  \Gamma(n+\tfrac{3}{2})\partial_n\\
&=\sum_{n=0}^\infty\sum_{k=0}^{n+1}
  \frac{\tilt_k}{\Gamma(k+\frac{1}{2})}
  \frac{\beta^{n-k+1}}{(n-k+1)!}
  \Gamma(n+\tfrac{3}{2})\partial_n.
\end{aligned}
\end{align}
In the last step we have used the string equation \eqref{eq:streq}.
By setting $m=n-k$ this is further rewritten as
\begin{align}
\begin{aligned}
&\sum_{\substack{m\ge -1,\ k\ge 0\\[.5ex] (m,k)\ne (-1,0)}}
  \frac{\tilt_k}{\Gamma(k+\frac{1}{2})}
  \frac{\beta^{m+1}}{(m+1)!}
  \Gamma(k+m+\tfrac{3}{2})\partial_{k+m}\\
&=\sum_{\substack{m\ge -1,\ k\ge 0\\[.5ex] (m,k)\ne (-1,0)}}
   \frac{(2k+2m+1)!!\beta^{m+1}}{(2k-1)!!(m+1)!2^{m+1}}
   \tilt_k\partial_{k+m}.
\end{aligned}
\end{align}
This is precisely the second term on the r.h.s.~of \eqref{eq:cL1}
and thus equal to the l.h.s.~of \eqref{eq:cLtdel}.

%%%%%%%%%%%%%%%%%%%%%%%%%%%%%%%%%%%%%%%%%%%%%%%%%%%%%%%%%%%%%%%%%%%%%%%%
\section{\mathversion{bold}Loop equation at finite $N$}\label{sec:finiteN}
%%%%%%%%%%%%%%%%%%%%%%%%%%%%%%%%%%%%%%%%%%%%%%%%%%%%%%%%%%%%%%%%%%%%%%%%

In this appendix we review the loop equation
of matrix model at finite $N$.
See e.g.~\cite{Ginsparg:1993is} for a review on this subject.

The loop equation of matrix model follows from the 
Schwinger-Dyson equation
\begin{equation}
\begin{aligned}
 \sum_{i,j=1}^N\int dM \frac{\del}{\del M_{ij}}\Biggl[\bigl(e^{\bt M}\bigr)_{ij}e^{-\frac{1}{\gs}\Tr V(M)+\int_0^\infty d\bt J(\bt)Z(\bt)}\Biggr]=0,
\end{aligned} 
\label{eq:SD}
\end{equation}
where $M$ is the $N\times N$ hermitian matrix and $Z(\bt)$ is defined by
\begin{equation}
\begin{aligned}
 Z(\bt)=\Tr e^{\bt M}.
\end{aligned} 
\end{equation}
$J(\bt)$ in \eqref{eq:SD} is the source for $Z(\bt)$.
Note that $Z(\bt)$ becomes the usual partition function $\Tr e^{-\bt H}$
under the identification $M=-H$.

Using the relations
\begin{equation}
\begin{aligned}
 \sum_{i,j=1}^N\frac{\del}{\del M_{ij}}\bigl(e^{\bt M}\bigr)_{ij}&=\int_0^\bt ds
\Tr e^{sM}\Tr e^{(\bt-s)M},\\
\sum_{i,j=1}^N\bigl(e^{\bt M}\bigr)_{ij}
\frac{\del}{\del M_{ij}}\Tr f(M)&=
\Tr\Bigl[f'(M)e^{\bt M}\Bigr],
\end{aligned} 
\end{equation}
\eqref{eq:SD} is written as
\begin{equation}
\begin{aligned}
 \frac{1}{\gs}\Biggl\bra\Tr\Bigl[V'(M)e^{\bt M}\Bigr]\Biggr\ket_J
=\int_0^\bt ds \Bigl\bra\Tr e^{sM}\Tr e^{(\bt-s)M}\Bigr\ket_J+
\int_0^\infty d\bt'\bt' J(\bt') \Bigl\bra\Tr e^{(\bt+\bt')M}\Bigr\ket_J,
\end{aligned}
\label{eq:loop-M} 
\end{equation}
where $\bra\cdots\ket_J$ is defined by
\begin{equation}
\begin{aligned}
 \bra\cdots\ket_J=\frac{\int dM(\cdots)e^{-\frac{1}{\gs}\Tr V(M)+\int_0^\infty d\bt J(\bt)Z(\bt)}}
{\int dM e^{-\frac{1}{\gs}\Tr V(M)}}.
\end{aligned} 
\end{equation}
Using the identity $\Tr V'(M)e^{\bt M}=V'(\del_\bt)\Tr e^{\bt M}$,
\eqref{eq:loop-M} is rewritten in terms of $Z(\bt)=\Tr e^{\bt M}$ as
\begin{equation}
\begin{aligned}
 \frac{1}{\gs}V'(\del_\bt)\big\bra Z(\bt)\big\ket_J
=\int_0^\bt ds\big\bra Z(s)Z(\bt-s)\big\ket_J+\int_0^\infty d\bt'\bt' J(\bt')
\big\bra Z(\bt+\bt')\big\ket_J.
\end{aligned} 
\label{eq:loop-J}
\end{equation}
By taking the derivative of the both sides of \eqref{eq:loop-J}
with respect to $J$ and setting $J=0$, we find the
loop equation for the multi-point functions of $Z(\bt)$. For instance,
by simply setting $J=0$ in \eqref{eq:loop-J} we find
\begin{equation}
\begin{aligned}
 \frac{1}{g_s}V'(\del_\bt)\big\bra Z(\bt)\big\ket
=\int_0^\bt ds\big\bra Z(s)Z(\bt-s)\big\ket.
\end{aligned} 
\end{equation}
By induction, one can show that the following loop equation is obtained from
\eqref{eq:loop-J}
\begin{equation}
\begin{aligned}
 &\frac{1}{g_s}V'(\del_\bt)Z_{n+1}(\bt,\bt_1,\cdots,\bt_n)\\
=&\int_0^\bt ds\left[Z_{n+2}(s,\bt-s,\bt_1,\cdots,\bt_n)
+\sum_{I\subset S}Z_{|I|+1}(s;\bt_I)Z_{|S-I|+1}(\bt-s;\bt_{S-I})\right]\\
&+\sum_{j=1}^n \bt_jZ_{n}\bigl(\{\bt_i+\cob_{ij}\bt\}_{i=1}^n\bigr),
\end{aligned} 
\label{eq:loop-finite}
\end{equation}
where $Z_n(\bt_1,\ldots,\bt_n)=\bra Z(\bt_1)\cdots Z(\bt_n)\ket_{\text{conn}}$
denotes the connected correlator and the definition of $I$ and $S$
is the same as in \eqref{eq:loopgen}. \eqref{eq:loop-finite} is the finite $N$
version of the loop equation; essentially it has the same form
as the loop equation \eqref{eq:loopgen} in the double-scaled matrix model
but the potential $V$ should be replaced by the effective potential $V_{\text{eff}}$
in the double-scaled version of the loop equation \eqref{eq:loopgen}. 
Note that the integral over $s$ in \eqref{eq:loop-finite} is finite 
and we do not have to subtract the genus-zero part of one-point function 
at finite $N$.

%%%%%%%%%%%%%%%%%%%%%%%%%%%%%%%%%%%%%%%%%%%%%%%%%%%%%%%%%%%%%%%%%%%%%%%%
\section{\mathversion{bold}Cut-and-join representation of $|V\ket$}\label{sec:V}
%%%%%%%%%%%%%%%%%%%%%%%%%%%%%%%%%%%%%%%%%%%%%%%%%%%%%%%%%%%%%%%%%%%%%%%%

In this appendix we review the derivation of $|V\ket$ in \eqref{eq:V-candj} 
\cite{Alexandrov:2010bn}.
The Virasoro constraint $\h{L}_m|V\ket=0$ is written as
\begin{equation}
\begin{aligned}
 \h{L}_m'|V\ket=\frac{1}{2\gs}\al_{2m+3}|V\ket,
\end{aligned} 
\label{eq:VL'}
\end{equation}
where $\h{L}_m'=\h{L}_m+\frac{1}{2\gs}\al_{2m+3}$.
Following \cite{Alexandrov:2010bn} we introduce the operator $D$ by
\begin{equation}
\begin{aligned}
 D=\hf\sum_{k=0}^\infty\al_{-2k-1}\al_{2k+1}.
\end{aligned} 
\end{equation}
One can show that
\begin{equation}
\begin{aligned}
{}[D,\h{L}_m']=-m\h{L}_m',\quad
[D,\al_{2k+1}]=-\Bigl(k+\hf\Bigr)\al_{2k+1}.
\end{aligned} 
\end{equation}
Then we expand $|V\ket$ as
\begin{equation}
\begin{aligned}
 |V\ket=\sum_{n=0}^\infty |V_n\ket,\quad
 D|V_n\ket=\lap_n|V_n\ket,
\end{aligned} 
\label{eq:Vn}
\end{equation}
where $|V_n\ket$ has the increasing weight of $D$
\begin{equation}
\begin{aligned}
 \lap_0<\lap_1<\lap_2<\cdots.
\end{aligned} 
\end{equation}
We assume that $|V_0\ket=|\Om\ket$ and $\lap_0=0$.
Plugging \eqref{eq:Vn} into \eqref{eq:VL'}  we find
\begin{equation}
\begin{aligned}
 \h{L}_m'|V_n\ket=\frac{1}{2\gs}\al_{2m+3}|V_{n+1}\ket.
\end{aligned} 
\end{equation}
From this we find the condition of the weight $\lap_n$
\begin{equation}
\begin{aligned}
 -m+\lap_n=-m-\frac{3}{2}+\lap_{n+1}.
\end{aligned} 
\end{equation}
This is solved as
\begin{equation}
\begin{aligned}
 \lap_n=\frac{3n}{2}.
\end{aligned} 
\end{equation}
Applying $\al_{-2m-3}$ to both sides of \eqref{eq:VL'} and summing over
$m\geq-1$ we find the recursion relation for $|V_n\ket$
\begin{equation}
\begin{aligned}
 \sum_{m=-1}^\infty \al_{-2m-3}\h{L}_m'|V_n\ket=\frac{1}{\gs}D|V_{n+1}\ket
=\frac{3(n+1)}{2\gs}|V_{n+1}\ket.
\end{aligned} 
\end{equation}
This is solved as
\begin{equation}
\begin{aligned}
 |V_n\ket=\frac{1}{n!}\left(\frac{2\gs}{3}\sum_{m=-1}^\infty \al_{-2m-3}\h{L}_m'\right)^n|\Om\ket.
\end{aligned} 
\label{eq:Vn-form}
\end{equation}
Finally, plugging \eqref{eq:Vn-form} into \eqref{eq:Vn}
we find our desired result \eqref{eq:V-candj}.
%%%%%%%%%%%%%%%%%%%%%%%%%%%%%%%%%%%%%%%%%%%%%%%%%%%%%%%%%%%%%%%%%%%%%%%%
\bibliography{paper}
\bibliographystyle{utphys}

\end{document}